\documentclass[%
 preprint, 
 amsmath,amssymb,
 aps, physrev,
]{revtex4-2}

\usepackage{graphicx}
\usepackage{dcolumn}
\usepackage{bm}
\graphicspath{{figures/}}
\usepackage[ruled,vlined,linesnumbered]{algorithm2e}
\SetAlgoCaptionLayout{centerline}
\SetAlgorithmName{Algorithm}{Algorithm}{List of Algorithms}
\usepackage{overpic}
\usepackage{tikz}
\tikzset{>=stealth}
\usepackage{xcolor}
\usepackage{hyperref}

\begin{document}

\preprint{}

\title{\textbf{Deep reinforcement learning based navigation of a jellyfish-like swimmer in flows with obstacles}}

\author{Yihao Chen}
 \affiliation{State Key Laboratory for Turbulence and Complex Systems, School of Mechanics and Engineering Science, Peking University, Beijing 100871, China}
\author{Yue Yang}%
 \email{yyg@pku.edu.cn.}
 \affiliation{State Key Laboratory for Turbulence and Complex Systems, School of Mechanics and Engineering Science, Peking University, Beijing 100871, China}
 \affiliation{HEDPS-CAPT, Peking University, Beijing 100871, China}

\date{\today}

\begin{abstract}
We develop a deep reinforcement learning framework for controlling a bio-inspired jellyfish swimmer to navigate complex fluid environments with obstacles. 
While existing methods often rely on kinematic and geometric states, a key challenge remains in achieving efficient obstacle avoidance under strong fluid-structure interactions and near-wall effects. 
We augment the agent's state representation within a soft actor-critic algorithm to include the real-time forces and torque experienced by the swimmer, providing direct mechanical feedback from vortex-wall interactions. 
This augmented state space enables the swimmer to perceive and interpret wall proximity and orientation through distinct hydrodynamic force signatures. 
We analyze how these force and torque patterns, generated by walls at different positions influence the swimmer's decision-making policy. 
Comparative experiments with a baseline model without force feedback demonstrate that the present one with force feedback achieves higher navigation efficiency in two-dimensional obstacle-avoidance tasks. 
The results show that explicit force feedback facilitates earlier, smoother maneuvers and enables the exploitation of wall effects for efficient turning behaviors. 
With an application to autonomous cave mapping, this work underscores the critical role of direct mechanical feedback in fluid environments and presents a physics-aware machine learning framework for advancing robust underwater exploration systems.
\end{abstract}

\maketitle

\section{\label{sec:intro}Introduction}
Deep reinforcement learning (DRL) \citep{richard2018RL,Viq2022reviewDRL} has emerged as an important tool for tackling complex control challenges in fluid dynamics, particularly those involving intricate fluid-structure interactions (FSI). 
While traditional computational fluid dynamics (CFD) methods often grapple with prohibitive computational costs and numerical complexities in real-time control scenarios, DRL offers a powerful data-driven alternative. 
It enables the autonomous learning of control policies that can adaptively stabilize, maneuver, or optimize fluid systems, demonstrating significant success in applications such as aerodynamic shape optimization~\citep{Duss2023shapeOpt,LU2025shape,Scavella2025DRLshapeOpt,Liu2024shapeOpt,Noda2023ShapeOpt}, drag reduction~\citep{Fan2020cyliner,han2022cylinder, Koumou2024drag,Lee2023DragReduct,Wang2024DragReduct,Tang2024cylinder,Cavallazzi2024dragReduct}, and active flow control~\citep{beckerehmck2020learning, Rodwell2023torque, Xie2023control,2025navigate,Wang2024Control,Zhu2022RLf,Wu2024control,Novati2017fishdrl}.

A demanding frontier for DRL lies in the control of soft-bodied underwater robots navigating confined, obstacle-laden spaces, such as underwater caves or wreck interiors \citep{Qu2023underwaterSoftrobot, Hasib2025underwarterSoftRobot}.
Soft robotic systems have operational safety advantage over traditional rigid robots in several aspects \citep{Liu2025Jet,Li2025SoftRob,Xu2024SoftRob}. 
First, their compliant nature provides inherent collision resilience. 
The energy from unintended impacts with obstacles is absorbed and dissipated through the robot's body, drastically reducing the risk of mechanical damage and ensuring mission continuity. 
Furthermore, this compliance translates into superior maneuverability. 
Soft robots can safely navigate through complex and confined environments, such as coral reefs or underwater ruins, without the need for forceful interaction that could damage both the robot and the surroundings. 

In these scenarios, agents face a dual challenge: the complex, often unintuitive dynamics of FSI inherent to flexible morphologies, and the profound hydrodynamic wall effects that dominate near boundaries \citep{HASIB2025softRob,Xue2023WallEff,Li2025WallEff, Terrington2024vortexWall,Zhang2025Vortex}. 
These effects can drastically alter a swimmer's dynamics, making precise, robust control difficult. 
This challenge is compounded by the limitations of conventional sensing.
Many current DRL strategies \citep{LI2024drl,ZHANG2024path,TANG2024path,WANG2024path,YANG2025path} for object motion control in fluid rely predominantly on kinematic or geometric state representations (e.g., position, velocity, proximity distances), often overlooking the direct physical force interactions between the compliant agent and its fluid environment as FSI does not have significant influence. 
For a soft robot, this limitation is critical, as its shape and the forces upon it are intrinsically linked, and traditional states may not be sufficient to capture the nuanced hydrodynamic cues essential for anticipating and mitigating collisions in tight spaces.

To bridge this gap, this work focuses on a fundamental aspect of physical interaction: the direct force and torque feedback experienced by a controlled soft body immersed in a fluid. 
We posit that for a compliant swimmer in a confined environment, incorporating these mechanical interaction signals as explicit state variables within the DRL framework provides essential, real-time information about boundary proximity and orientation, enabling more informed and anticipatory decision-making for navigation and control.

Specifically, we investigate this hypothesis using a bio-inspired, DRL-controlled 2D jellyfish swimmer \citep{hoover2015jelly} navigating obstacle-laden environments. 
Bio-inspired design shows advanced performance in various application \citep{Liu2025Bio,Novati2017fishdrl}.
Jellyfish exhibit high locomotion efficiency through body deformation and FSI \citep{Brad2013passive,Costello2021efficent}. 
Their swimming mechanics, characterized by energy-efficient propulsion \citep{Nicole2020propulsion, KB2012Modeling, FJ2018Thrust, Matharu2022jellyZ} and morphological simplicity, enable effective maneuverability \citep{hoover2015jelly}. 
These traits make jellyfish-inspired robots promising for ocean exploration and ecosystem research \citep{Anne2019ocean1, Reaka-Kudla2001Pearl, Tim2019oceantrend}.
Building upon prior work \citep{chen2024DRL}, our core contribution lies in augmenting the agent's state representation to include the forces and torque exerted by the surrounding fluid on the jellyfish body.
We analyze how the distinct force or torque signatures generated by walls at different relative positions influence the swimmer. 
Through comparative numerical experiments between the previous agent \citep{chen2024DRL} and our augmented agent on a simple single-obstacle target pursuit task, we demonstrate that explicit force feedback enhances navigation efficiency, enabling faster target acquisition and obstacle avoidance. 
This finding contributes to advancing DRL strategies for robust underwater navigation in challenging, real-world applications such as cave exploration and infrastructure inspection.

\section{\label{sec:DRL}Deep Reinforcement Learning}
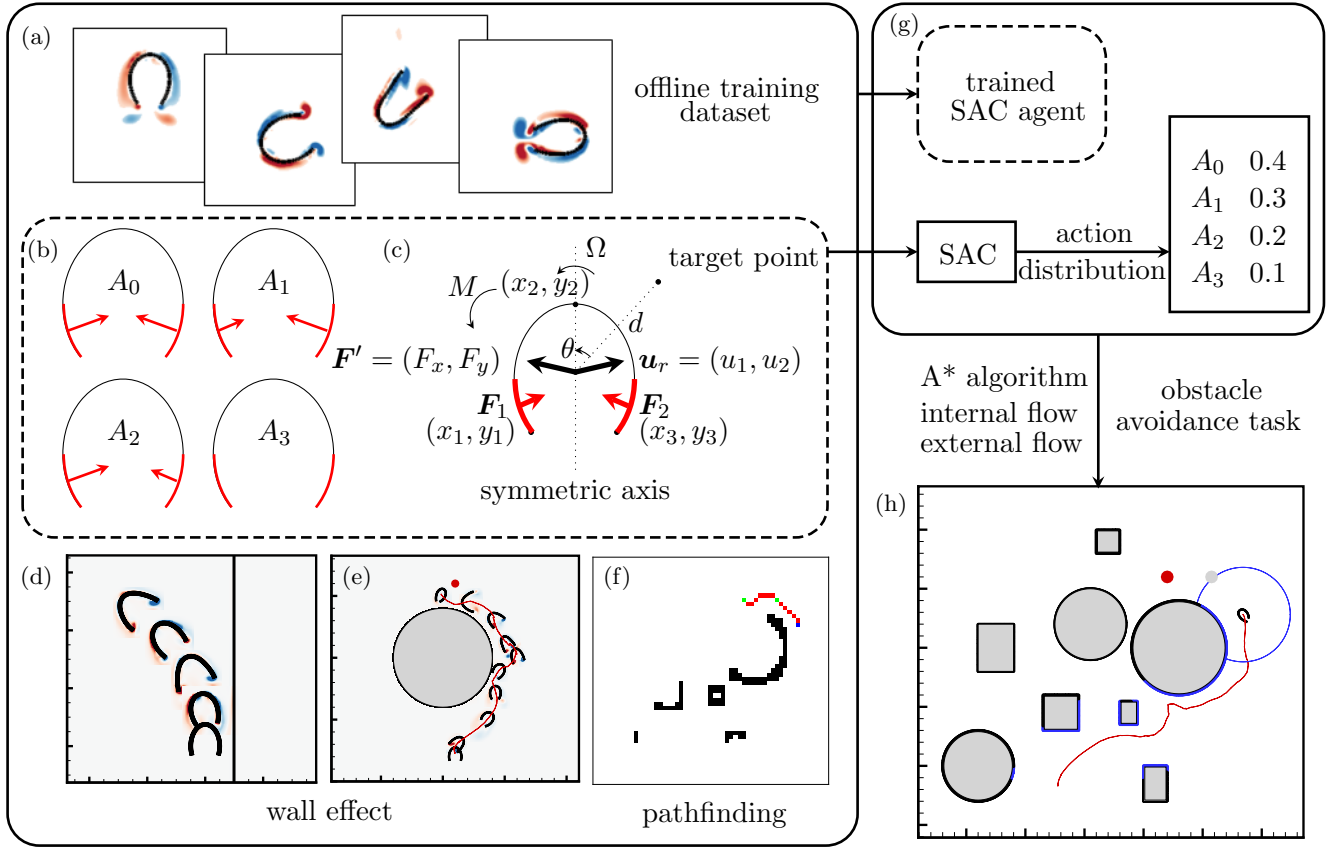
\begin{figure}
	\centering
	\begin{overpic}[scale=0.42]{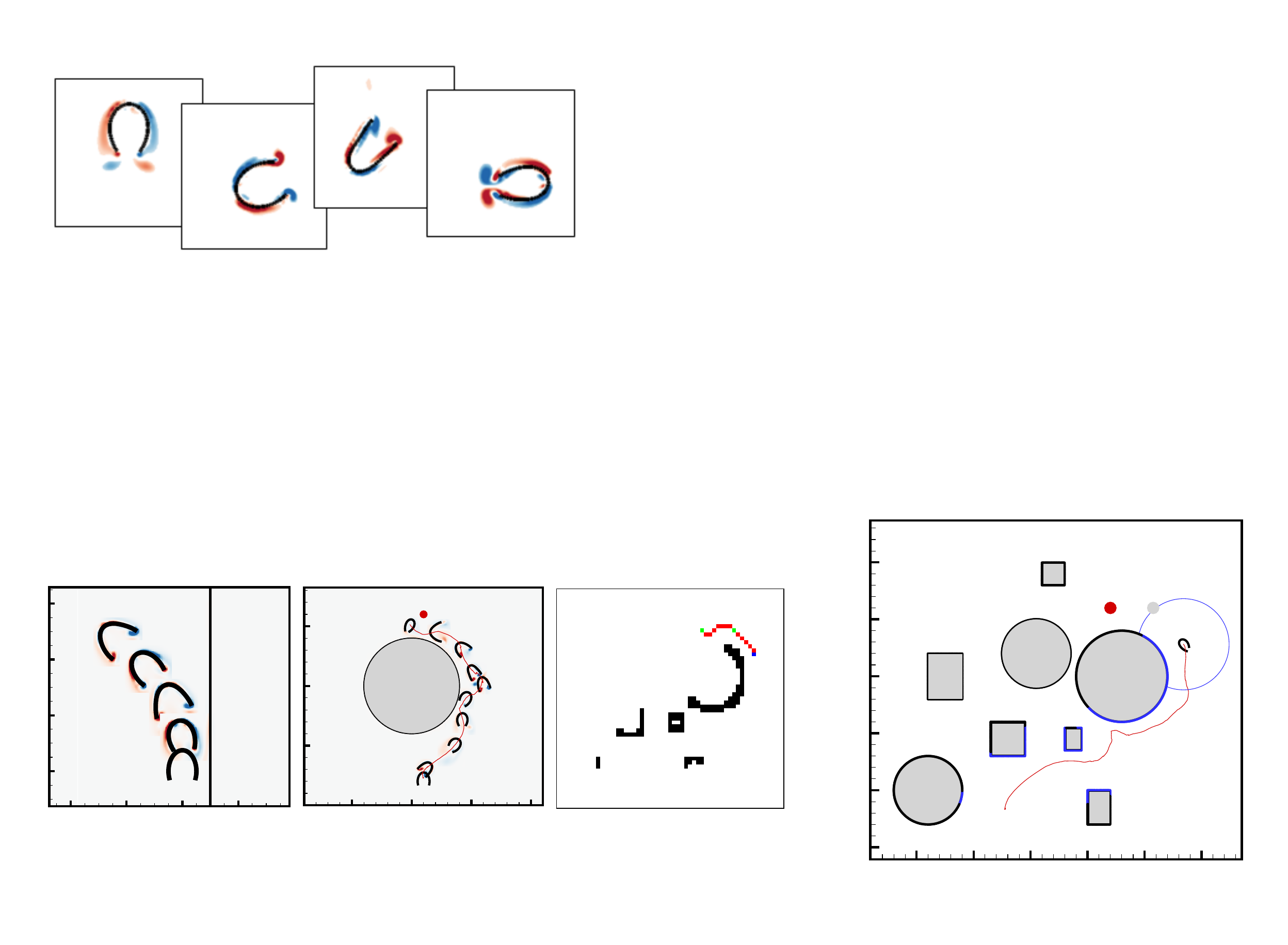}
     \put(-17,66){%
			\begin{minipage}{0.4\textwidth} %
				\fontsize{9}{14}\selectfont %
				(a)
			\end{minipage}
		}
      \put(-16.5,50){%
			\begin{minipage}{0.4\textwidth} %
				\fontsize{9}{14}\selectfont %
				(b)
			\end{minipage}
		}
      \put(10,50){%
			\begin{minipage}{0.4\textwidth} %
				\fontsize{9}{14}\selectfont %
				(c)
			\end{minipage}
		}
      \put(-17,25.5){%
			\begin{minipage}{0.4\textwidth} %
				\fontsize{9}{14}\selectfont %
				(d)
			\end{minipage}
		}
  \put(7,25.5){%
			\begin{minipage}{0.4\textwidth} %
				\fontsize{9}{14}\selectfont %
				(e)
			\end{minipage}
		}
      \put(26.5,25.5){%
			\begin{minipage}{0.4\textwidth} %
				\fontsize{9}{14}\selectfont %
				(f)
			\end{minipage}
		}
      \put(48,67){%
			\begin{minipage}{0.4\textwidth} %
				\fontsize{9}{14}\selectfont %
				(g)
			\end{minipage}
		}
      \put(47,31){%
			\begin{minipage}{0.4\textwidth} %
				\fontsize{9}{14}\selectfont %
				(h)
			\end{minipage}
		}
		\put(45,37){%
			\begin{tikzpicture}[overlay,scale=1]
        \draw (0,0) arc (-45:225:0.8 and 1);
        \draw [dotted] (-0.55,-0.5)--(-0.55,2.5);
        \draw [dotted] (-0.55,0.8)--(0.55,2);
        \draw (-0.35,1)[->] arc (45:90:0.3);
        \node[above right] at (-0.88,0.85){$\theta$};
        \draw (-0.3,2.1)[->] arc (45:135:0.35);
        \node[above right] at (-0.55,2.2){$\Omega$};
        \node[above right] at (0.55,2){target point};
        \node[below] at (-0.55,-0.5){symmetric axis};
        \node[right] at (0.15,0){$(x_{3},y_{3})$};
        \node[left] at (-1.2,0){$(x_{1},y_{1})$};
        \node[above] at (-0.95,1.65){$(x_{2},y_{2})$};
        \filldraw (-1.14,0) circle (0.03);
        \filldraw (0,0) circle (0.03);
        \filldraw (-0.55,1.7) circle (0.03);
        \filldraw (0.55,2) circle (0.03);
        \filldraw (-0.55,0.8) circle (0.03);
        \node[above] at (0.25,1.2){$d$};
        \draw[line width=2pt, color=red] (0,0) arc (-45:0:0.8 and 1);
        \draw[line width=2pt, color=red] (-1.12,0) arc (225:180:0.8 and 1);
        \draw[line width=2pt, color=red][->](-1.3,0.35)--(-0.95,0.5);
        \draw[line width=2pt, color=red][->](0.15,0.35)--(-0.2,0.5);
        \node[left] at (-1.3,0.35){$\boldsymbol{F}_{1}$};
        \node[right] at (0.15,0.35){$\boldsymbol{F}_{2}$};
        \draw[line width=2pt][->](-0.55,0.8)--(0.1,0.95);
        \node[right] at (0.15,0.95) {$\boldsymbol{u}_r=(u_{1},u_{2})$};
        \draw[line width=2pt][->](-0.55,0.8)--(-1.2,0.95);
        \node[left] at (-1.40,0.95) {$\boldsymbol{F}^\prime=(F_{x},F_{y})$};
        \draw (-1.6,1.9)[->] arc (90:210:0.35);
        \node[left] at (-1.7,1.95){$M$};
        \draw (-6,-1) arc (-45:225:0.8 and 1);
        \draw[line width=1pt, color=red] (-6,-1) arc (-45:0:0.8 and 1);
        \draw[line width=1pt, color=red] (-7.12,-1) arc (225:180:0.8 and 1);
        \draw[line width=1pt, color=red][->](-7.3,-0.65)--(-6.75,-0.45);
        \draw[line width=1pt, color=red][->](-5.85,-0.65)--(-6.2,-0.5);
        \node[above] at (-6.55,-0.3){$A_2$};
        \draw (-4,-1) arc (-45:225:0.8 and 1);
        \draw[line width=1pt, color=red] (-4,-1) arc (-45:0:0.8 and 1);
        \draw[line width=1pt, color=red] (-5.12,-1) arc (225:180:0.8 and 1);
        \node[above] at (-4.55,-0.3){$A_3$};
        \draw (-6,1) arc (-45:225:0.8 and 1);
        \draw[line width=1pt, color=red] (-6,1) arc (-45:0:0.8 and 1);
        \draw[line width=1pt, color=red] (-7.12,1) arc (225:180:0.8 and 1);
        \draw[line width=1pt, color=red][->](-7.3,1.35)--(-6.75,1.55);
        \draw[line width=1pt, color=red][->](-5.85,1.35)--(-6.4,1.55);
        \node[above] at (-6.55,1.7){$A_0$};
        \draw (-4,1) arc (-45:225:0.8 and 1);
        \draw[line width=1pt, color=red] (-4,1) arc (-45:0:0.8 and 1);
        \draw[line width=1pt, color=red] (-5.12,1) arc (225:180:0.8 and 1);
        \draw[line width=1pt, color=red][->](-5.3,1.35)--(-4.95,1.5);
        \draw[line width=1pt, color=red][->](-3.85,1.35)--(-4.4,1.55);
        \node[above] at (-4.55,1.7){$A_1$};
        \draw[dash pattern=on 4pt off 2pt,rounded corners=12pt,line width=1pt] (-7.9,-1.4) rectangle (2.8,2.85);
        \draw[rounded corners=12pt,line width=1pt] (-8.1,-5.5) rectangle (3.2,5.7);
        \draw[line width=1pt][->](3.2,4.5)--(4.0,4.5);
        \node[above right] at (0.1,4.25){offline training};
        \node[above right] at (0.7,4.0){dataset};
        \node[above right] at (-4.8,-5.3){wall effect};
        \node[above right] at (0.2,-5.4){pathfinding};
        \draw[dash pattern=on 4pt off 2pt,rounded corners=12pt,line width=1pt] (4.0,3.6) rectangle (6.5,5.4);
        \node[above right] at (4.5,4.4){trained};
        \node[above right] at (4.3,3.95){SAC agent};
        \draw[line width=1pt][->](2.8,2.4)--(4.0,2.4);
        \draw[line width=1pt] (4.0,2) rectangle (5.3,2.8);
        \node[above right] at (4.15,2.1){SAC};
        \draw[line width=1pt][->](5.3,2.4)--(7.35,2.4);
        \draw[line width=1pt] (7.35,1.6) rectangle (9.2,4.2);
        \node[above right] at (5.7,2.4){action};
        \node[above right] at (5.2,1.9){distribution};
        \node[above right] at (7.5,1.8){$A_3\,\,\,\,\,0.1$};
        \node[above right] at (7.5,2.3){$A_2\,\,\,\,\,0.2$};
        \node[above right] at (7.5,2.8){$A_1\,\,\,\,\,0.3$};
        \node[above right] at (7.5,3.3){$A_0\,\,\,\,\,0.4$};
        \draw[rounded corners=12pt,line width=1pt] (3.4,1.3) rectangle (9.4,5.7);
        \draw[line width=1pt][->](6.4,1.3)--(6.4,-0.75);
        \node[above right] at (7.1,0.3){obstacle};
        \node[above right] at (6.5,-0.1){avoidance task};
        \node[above right] at (3.9,0.4){A* algorithm};
        \node[above right] at (3.9,0.0){internal flow};
        \node[above right] at (3.9,-0.4){external flow};
\end{tikzpicture}
		}
\end{overpic}
	\caption{Diagram for the overall workflow. 
         (a) Data obtained from multiple simulations are used for offline training. 
         (b) Action space with four actions $A_i$, $i=0,1,2,3$, representing typical jellyfish actions (from left to right): symmetric forces on the two sides, larger force on the right side, larger force on the left side, and no force. 
         (c) Geometry and state of the jellyfish-like swimmer. The red parts indicate where the forces are applied.
         (d) The swimmer choosing the moving forward action deviates with the existence of a side wall.
         (e) The swimmer with SAC agent on a simple obstacle avoidance task.
         (f) A pathfinding algorithm is performed in descretized domain based on detected boundaries. 
         (g) The SAC module receives the state vector and outputs a probability distribution for the actions. The action is chosen with this distribution. 
         (h) The swimmer senses the environment and uses the detected boundaries (in blue) to find a path in a map like (f). A pilot point is then calculated to guide the swimmer toward the target.}\label{fig:bigDiag}
\end{figure}

The overall workflow is sketched in Fig.~\ref{fig:bigDiag}, which is extended from that in Ref.~\citep{chen2024DRL} by incorporating force and torque feedback.
Multiple simulations were run to collect data of jellyfish-like swimming in Fig.~\ref{fig:bigDiag}(a). The dataset was then used for offline training of DRL agent in Fig.~\ref{fig:bigDiag}(f). 
The trained agent is employed for navigation in unknown environment with obstacles in Fig.~\ref{fig:bigDiag}(g). 
Figures~\ref{fig:bigDiag}(b,c) show the jellyfish's state, action space, and decision making process, which are later detailed in Sections~\ref{sec:setup} and \ref{sec:training}.
Figures~\ref{fig:bigDiag}(d,e) illustrate the wall effect and how the agent exploits it and finishes the task in short time. 
The code and dataset of our study are available online \citep{chen2025code}.

\subsection{\label{sec:data_pre}Data preparation}
The 2D flow data for the DRL of jellyfish-like swimming are obtained from numerical simulations. The immersed boundary method \citep{peskin2002immersed,Tong2021IB} is used to treat the fluid-solid coupling at the moving boundary.
A unit density, incompressible flow is governed by the Navier--Stokes equations
\begin{eqnarray}
	\frac { \partial \boldsymbol { u } } { \partial t } + \boldsymbol { u } \cdot \nabla \boldsymbol { u } &=& - \nabla p + \nu \nabla ^ { 2 } \boldsymbol { u } + \boldsymbol { f },\label{eq:ns_eqn} \\
	\nabla \cdot  \boldsymbol { u }&=&0,\label{eq:consective_eqn}
\end{eqnarray}
where $\boldsymbol { u }$, $p$, $\nu$ and $\boldsymbol {f}$ denote the velocity, pressure, kinematic viscosity and body force exerted by a jellyfish-like swimmer.

The immersed boundary is represented by Lagrangian markers.
A regularised delta function $\delta_h$ \citep{peskin2002immersed} is employed to interpolate and spread $\boldsymbol {f}$ between Eulerian and Lagrangian points, whose coordinates are denoted by $\boldsymbol {x}$ and $\boldsymbol {X}$, respectively.
The Eulerian force in Eq.~\eqref{eq:ns_eqn} is calculated by
\begin{equation}\label{eq: eulerian_lagrangian_force}
	\boldsymbol { f } ( \boldsymbol { x }) = \int_{S}  \boldsymbol { F } ( \boldsymbol { X }) \delta_h ( \boldsymbol { x } - \boldsymbol { X } ) d \boldsymbol{X},
\end{equation}
where $\boldsymbol{F}$ denotes the Lagrangian force at $\boldsymbol{X}$, and $S$ the domain of the immersed boundary.
The non-slip condition is satisfied by exerting $\boldsymbol{F}$ on the immersed boundary.
The velocity on the immersed boundary satisfies
\begin{equation}\label{eq:no_slip_vel}
	\int _ { \mathcal{D} } \boldsymbol { u } ( \boldsymbol { x }  ) \delta_h (\boldsymbol { x } - \boldsymbol { X }) d \boldsymbol { x } = \boldsymbol { U } _ { b } ( \boldsymbol { X }),
\end{equation}
where $\mathcal{D}$ denotes the entire fluid domain and $\boldsymbol { U } _ { b }$ the velocity at Lagrangian points.

The simulations were conducted using the code IBAMR \citep{GRIFFITH2007IB}, which is a distributed-memory parallel implementation of the immersed boundary method with adaptive mesh refinement for the Cartesian grid. 
The total number of grid points is in the order of $10^5$. 
We have conducted convergence tests by doubling the grid resolution to ensure that the target tracking trajectory converges.
We use the same data collecting and processing method as described in \citep{chen2024DRL}.

\subsection{\label{sec:SAC}Soft actor-critic method}
Soft actor-critic (SAC) \citep{haarnoja2019SAC} is a model-free, off-policy actor-critic reinforcement learning algorithm. 
It is designed to achieve a balance between exploration and exploitation by incorporating the principles of maximum entropy into reinforcement learning. 
This method not only maximizes the expected cumulative rewards but also encourages the policy to explore a wide range of actions, leading to more robust and stable learning.
SAC extends traditional reinforcement learning objectives by adding an entropy term to the reward function. 
The modified objective aims to maximize the expected cumulative reward over a policy $\pi$ as 
\begin{equation}\label{eq: SAC_value}
	J(\pi)=\mathbb{E}_{\pi}\left(\sum_{t=0}^{N_t} r_t + \alpha \mathcal{H}(\pi(\cdot|s_t)) \right),
\end{equation} 
where $r_t$ is the reward received at time step $t$, $\pi(\cdot|s_t)$ the policy, which is the probability distribution of choosing each action at state $s_t$; $\mathcal{H}(\pi(\cdot|s_t))$ represents the entropy of the policy, and temperature $\alpha$ controls the trade-off between maximizing rewards and maintaining high entropy. 
High entropy promotes exploration by encouraging diverse action sampling.

Unlike on-policy algorithms such as proximal policy optimization (PPO) \citep{schulman2017PPO}, SAC can be off-policy, meaning it can reuse past experiences stored in a replay buffer without interacting with the environment constantly. 
This makes SAC more sample-efficient and suitable for tasks where data collection is expensive or time-consuming.

\subsection{\label{sec:setup}Case setup}
The jellyfish-like locomotion is enabled by applying a pair of sinusoidal forces to the swimmer's tips (marked in red), as illustrated in Fig.~\ref{fig:bigDiag}(c). 
The force density (force per unit length)
\begin{equation}\label{eq:force_density}
	\boldsymbol{F}_{i}= F_i\,\mathrm{sin}(2\pi ft)\boldsymbol{\tau}_{i},~~~i=1,2,
\end{equation}
acting on the left and right halves of the swimmer are denoted as $\boldsymbol{F}_1$ and $\boldsymbol{F}_2$, respectively, with $F_i = |\boldsymbol{F}_{i}|$ and the unit tangent vector $\boldsymbol{\tau}_{i}$ on the swimmer's tips that point inward.
Note that $F_1$ and $F_2$ in Eq.~\eqref{eq:force_density} can be different, while they have the same frequency $f$ and zero initial phase.

We divide a period $t = 0 \sim T$ of the sinusoidal force pair $(F_1,F_2)$ applied to the swimmer into four quarters, with $T=1/f$. 
The force pair is non-dimensionalized by the reference force density 1 $\mathrm{N}/\mathrm{m}$.
During each quarter, a SAC agent is employed to process the current state and produce $(F_1,F_2)$.
In order to reduce the training complexity, we restrict the choices for $(F_1,F_2)$ to $(0.003,0.003)$ for symmetric force application (moving forward), $(0.001,0.003)$ for larger force on right-side muscle (turning right), $(0.003,0.001)$ for larger force on left-side muscle (turning left), and $(0,0)$ for no force applied (drifting), labeled as actions $A_i$ with $i=0$, 1, 2, and 3, respectively.

In our validation tests, the obstacles are unknown to the swimmer. 
Instead, the swimmer has a sensing range of $\mathcal{R}=4 d_0$, where $d_0$ is the diameter of the swimmer. 
The boundary Lagrangian points of obstacles are known to the swimmer only when they are within this range and not blocked by other obstacles.
A fixed target point defines the navigation objective. 
Throughout the maneuver, an auxiliary target point (referred to as ``pilot'' below)
Specifically, at each navigation step, the A* algorithm computes an optimal path from the swimmer's current position to the target, utilizing the locally known obstacle map (see Appendix~\ref{sec:Astar} for details).
Then, a pilot is selected along this computed path. 
Its distance from the swimmer is set approximately equal to the sensing range $\mathcal{R}$.
This positioning leverages the swimmer's maximum perception capability to provide timely guidance while avoiding obstacles.

The parameters for the swimmer are listed in Table \ref{table:para}.
The four actions patterns are shown in Fig.~\ref{fig:bigDiag}(b) and the same action regulation in \citep{chen2024DRL} is also adopted. 

\begin{table}
\caption{\label{table:para} Parameters in the simulation for jellyfish-like swimming.}
\begin{ruledtabular}
\begin{tabular}{lcr}
\textrm{Parameter}&
\textrm{Symbol}&
\textrm{Value}\\
\colrule
Swimmer diameter  & $d_{0}$ & 0.1 \\
Domain length & $W$  & 1.6\\
Frequency & $f$  & 0.5\\
Period & $T=1/f$ & 2\\
Kinematic viscosity & $\nu$ & $5\times10^{-5}$ \\
Characteristic velocity & $U=fd_0$ & 0.05\\
Reynolds number & $Re = Ul/\nu$ & 100\\
Discretization length & $\Delta x$ & $\frac{d_0}{2}$\\
Time step size & $\Delta t$ & $\frac{T}{4}$\\
Sensing range & $\mathcal{R}$ & $4d_0$\\
\end{tabular}
\end{ruledtabular}
\end{table}

\subsection{\label{sec:training}Training}
The state of the swimmer is characterized as a 15-dimensional vector
\begin{equation}\label{eq:state_vec}
	\boldsymbol{s}=(x_{1},y_{1},x_{2},y_{2},x_{3},y_{3},u_{1},u_{2},d,\theta,\Omega,F_x,F_y,M,n),
\end{equation}
where $(x_{i},y_{i}),~i=1,2,3$ denote the coordinates of the swimmer's left, middle, and right endpoints referenced to its mass center, $\boldsymbol{u}_r = (u_{1},u_{2})$ the velocity of the mass center relative to target, $d$ the distance between the mass center and target, $\theta$ the angle between the swimmer's symmetric axis and the line segment connecting the mass center and target, $\Omega$ the angular velocity of the symmetric axis, $(F_x,F_y)$ the force exerted on the swimmer, $M$ the torque refrenced to the swimmer's mass center, and $n=0,1,2,3$ indicates the quarter of period $T$.
Compared with the state vector in Ref.~\citep{chen2024DRL}, $F_X,F_y$, and $M$ are newly added. 
These force and torque signatures are highly correlated with wall proximity and orientation, which will be discussed in Section~\ref{sec:dynamics}. 

The actions $A_i,i=0,1,2,3$ illustrated in Fig.~\ref{fig:bigDiag}(b) correspond to symmetric, asymmetric, and no force applied to two halves of the swimmer. 
The reward function~\citep{chen2024DRL} 
\begin{equation}\label{eq:reward}
	r(\boldsymbol{s},a)=-\mathrm{min}(\theta^{2},3)+\mathcal A\,\mathrm{clip}(v,-0.1,0.1)-\mathcal B d
\end{equation}
is adopted, where $v$ is the projection of mass center's velocity onto the line segment connecting the mass center and target point, and constants $\mathcal A$ and $\mathcal B$ are tuned during the training; the clip function $\mathrm{clip}(x,b,c)$ returns $x$ if $b\le x\le c$, $b$ if $x<b$, and $c$ if $x>c$.
The training dataset is collected from CFD simulations with random actions and data augmentation method~\citep{chen2024DRL}. 
More details on training can be found in Appendix~\ref{sec:train_detail}.

\section{\label{sec:dynamics}Wall effects on jellyfish-like swimming} 
\subsection{\label{sec:wall}Vortex-wall interaction}
\begin{figure}
	\centering
	\begin{overpic}[scale=0.6]{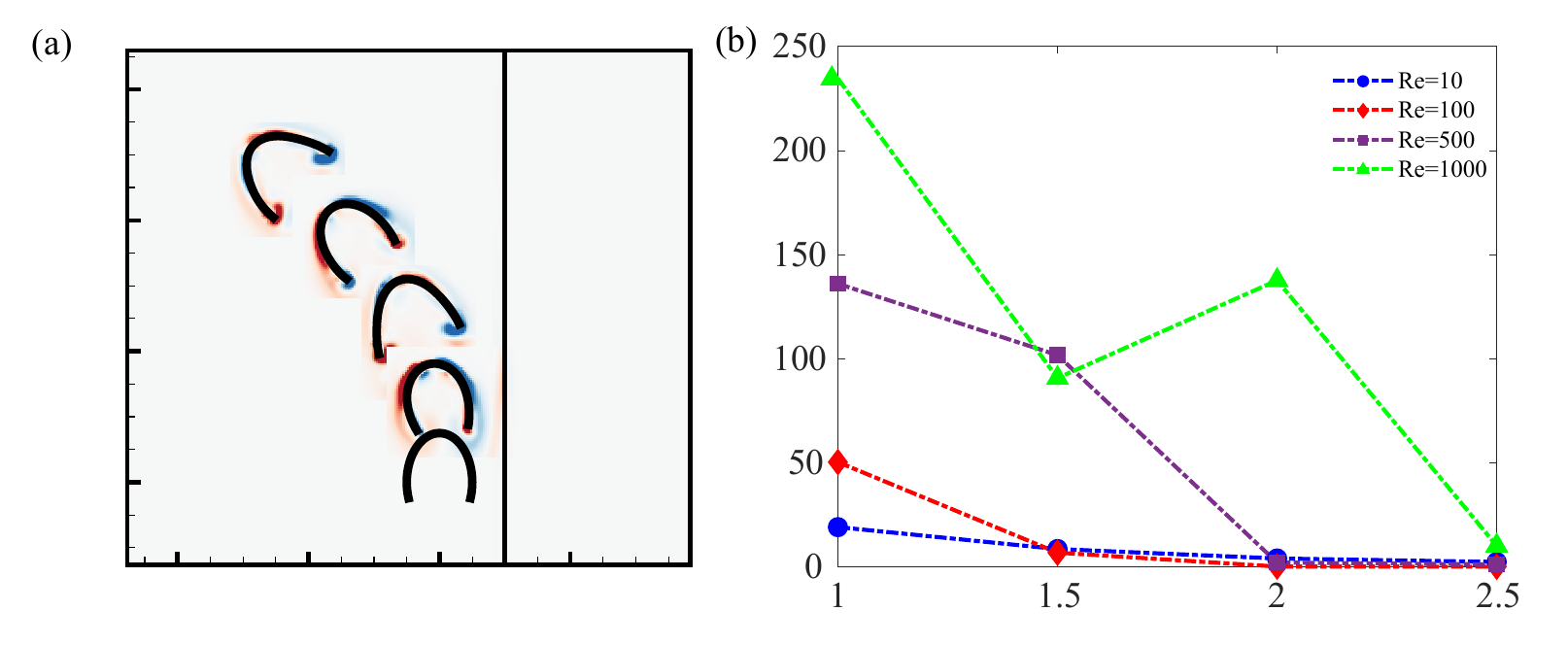}
		\put(55,0){%
			\begin{minipage}{0.4\textwidth} %
				\fontsize{12}{14}\selectfont %
				$d_w/d_0$ %
			\end{minipage}
		}
      \put(27,23){%
			\begin{minipage}{0.4\textwidth} %
				\fontsize{12}{14}\selectfont %
				$\varphi/{}^\circ$ %
			\end{minipage}
		}
      \put(-3.5,34.5){%
			\begin{minipage}{0.4\textwidth} %
				\fontsize{12}{14}\selectfont %
				$\varphi$ %
			\end{minipage}
		}
      \put(10,33.5){%
			\begin{minipage}{0.4\textwidth} %
				\fontsize{12}{14}\selectfont %
				$d_w$ %
			\end{minipage}
		}
      \put(15,30){%
			\begin{tikzpicture}[overlay,scale=1]
            \draw (0.5,0)--(0.5,1.1);
            \draw (0.5,0)--(-0.3,0.96);
            \draw (0.5,0.6)[->] arc (90:150:0.35);
            \draw (2.13,-3.5)--(2.13,1);
            \draw [->](1.53,0.7)--(2.13,0.7);
            \draw [->](3.4,0.7)--(2.8,0.7);
            \end{tikzpicture}
		}
\end{overpic}
	\caption{(a) Trajectory of the swimmer's forward motion with surrouding vorticity where the action $A_0$ is always chosen, with a side wall, where $\varphi$ and $d_r$ are the deflection angle of the swimmer's symmetric axis and distance from the swimmer's mass center to wall, respectively. (b) Angle deviation of swimmer swimming forward over duration of $t/T=10$ with different Reynolds numbers and initial distances to the side wall. }\label{fig:wall_perturb_traj} 
\end{figure}
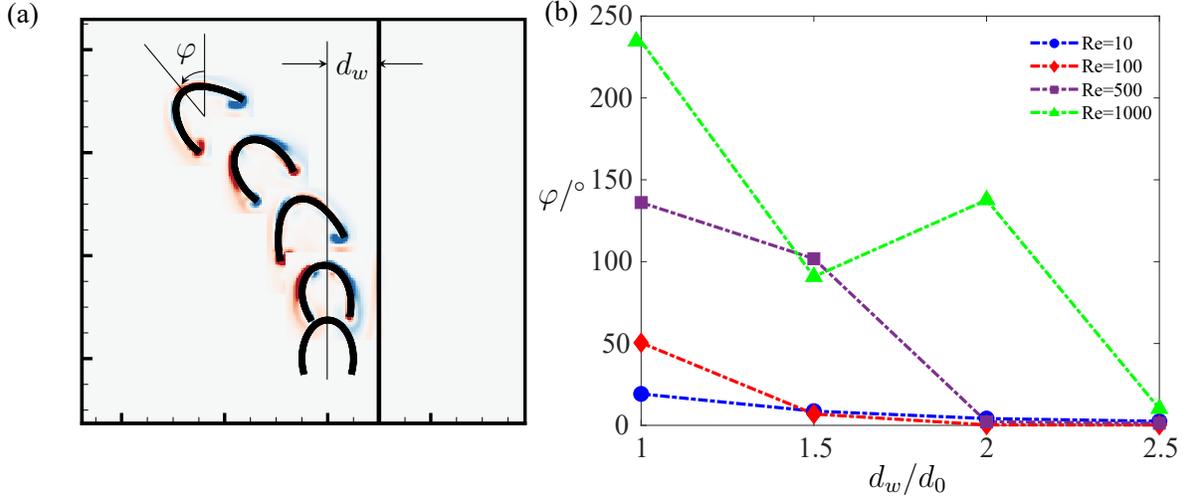

The presence of a wall exerts a significant influence on the surrounding flow field of a swimmer. 
Specifically, the wall generates a reflection flow field which affects the swimmer's trajectory. 
For example, as shown in Fig.~\ref{fig:wall_perturb_traj}(a), the swimming trajectory of a forward-swimming swimmer is strongly deflected when a side wall is introduced. 
We conducted simulations for multiple cases with varying initial wall distances $d_w$ between the swimmer's mass center and the side wall for different Reynolds numbers ($Re$). 
Figure~\ref{fig:wall_perturb_traj}(b) plots the deflection angle $\varphi$ of the swimmer's symmetric axis over a duration of $t/T = 10$. 

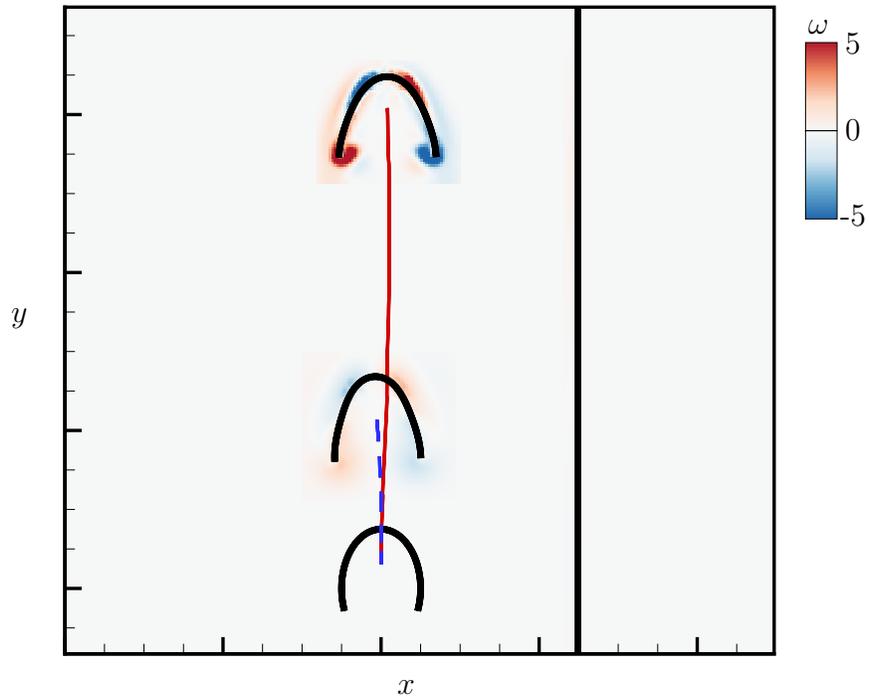
\begin{figure}
	\centering
	\begin{overpic}[scale=0.6]{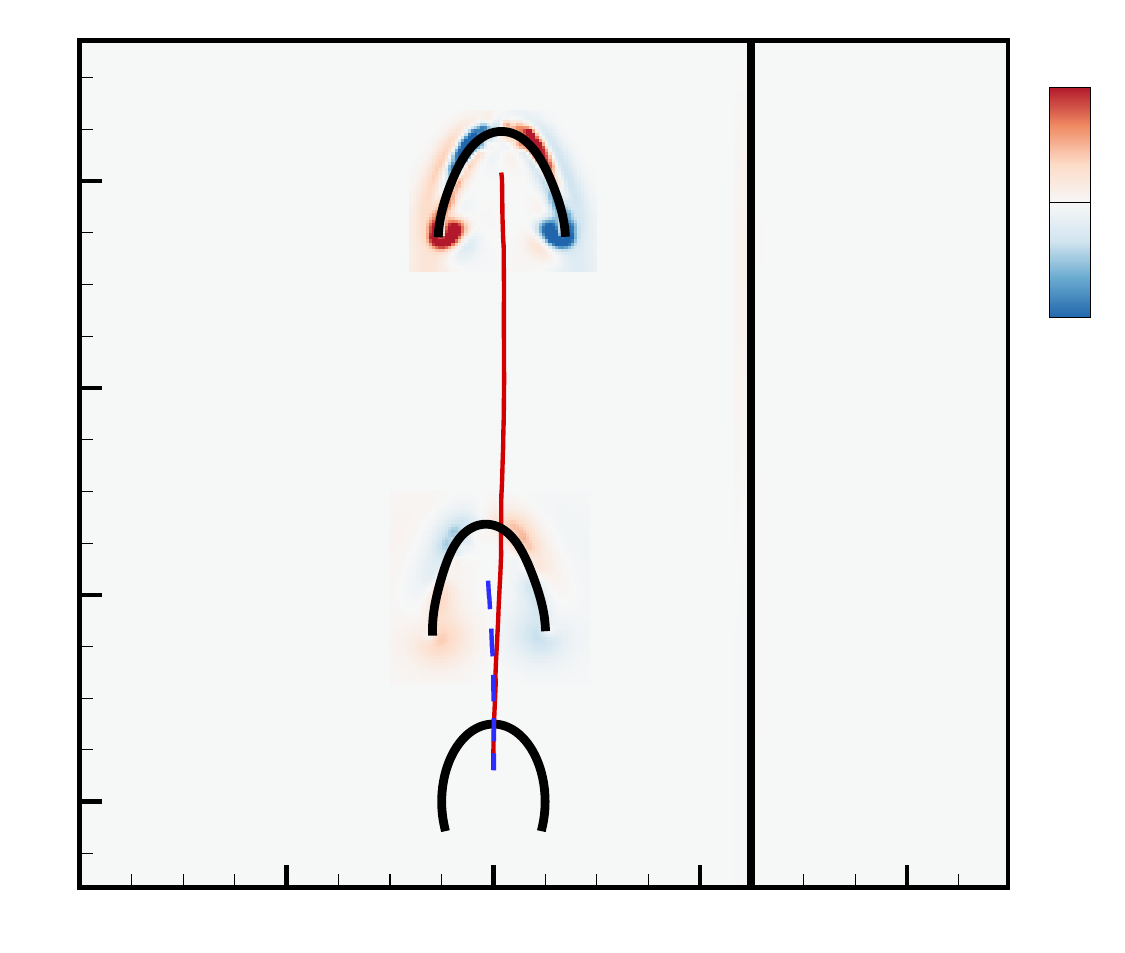}
		\put(18,1){%
			\begin{minipage}{0.4\textwidth} %
				\fontsize{12}{14}\selectfont %
				$x$ %
			\end{minipage}
		}
      \put(-27,44){%
			\begin{minipage}{0.4\textwidth} %
				\fontsize{12}{14}\selectfont %
				$y$ %
			\end{minipage}
		}
  \put(66,78){%
			\begin{minipage}{0.4\textwidth} %
				\fontsize{12}{14}\selectfont %
				$\omega$ %
			\end{minipage}
		}
  \put(70,76){%
			\begin{minipage}{0.4\textwidth} %
				\fontsize{12}{14}\selectfont %
				5 
			\end{minipage}
		}
  \put(70,66){%
			\begin{minipage}{0.4\textwidth} %
				\fontsize{12}{14}\selectfont %
				0 
			\end{minipage}
		}
  \put(70,56){%
			\begin{minipage}{0.4\textwidth} %
				\fontsize{12}{14}\selectfont %
				-5 
			\end{minipage}
		}
\end{overpic}
	\caption{Trajectory and vorticity field of the swimmer forward motion with a side wall at $d_w/d_0=2.5$ over duration of $t/T=10$ for $Re=10$ (blue dashed line) and $100$ (red solid line). 
 }\label{fig:lowRe_comp}
\end{figure}

The result reveals a distinct and non-monotonic influence of the wall proximity at different $Re$.
At low and moderate $Re$, the deflection angle exhibits a predictable monotonic decay as the wall distance increases. 
This decay is markedly slower in the low-$Re$ case, a consequence of the long-range nature of viscous interactions governed by the elliptic Stokes equations at the low-$Re$ limit. 
In this regime, the no-slip condition at both the body and wall surfaces creates a globally coupled flow field where the viscous stress diffuses radially, ensuring that the wall's influence is felt strongly even at larger wall distances, resulting in a persistent deflection force.
Figure~\ref{fig:lowRe_comp} compares the trajectories of the swimmer at $Re=10$ and $100$ with $d_w/d_0=2.5$ over duration of $t/T=10$.
The swimmer at $Re=10$ has a larger deflection angle and moves at a significantly lower speed than those at $Re=100$ where it has almost zero deflection angle and minor displacement in the wall-normal direction. 

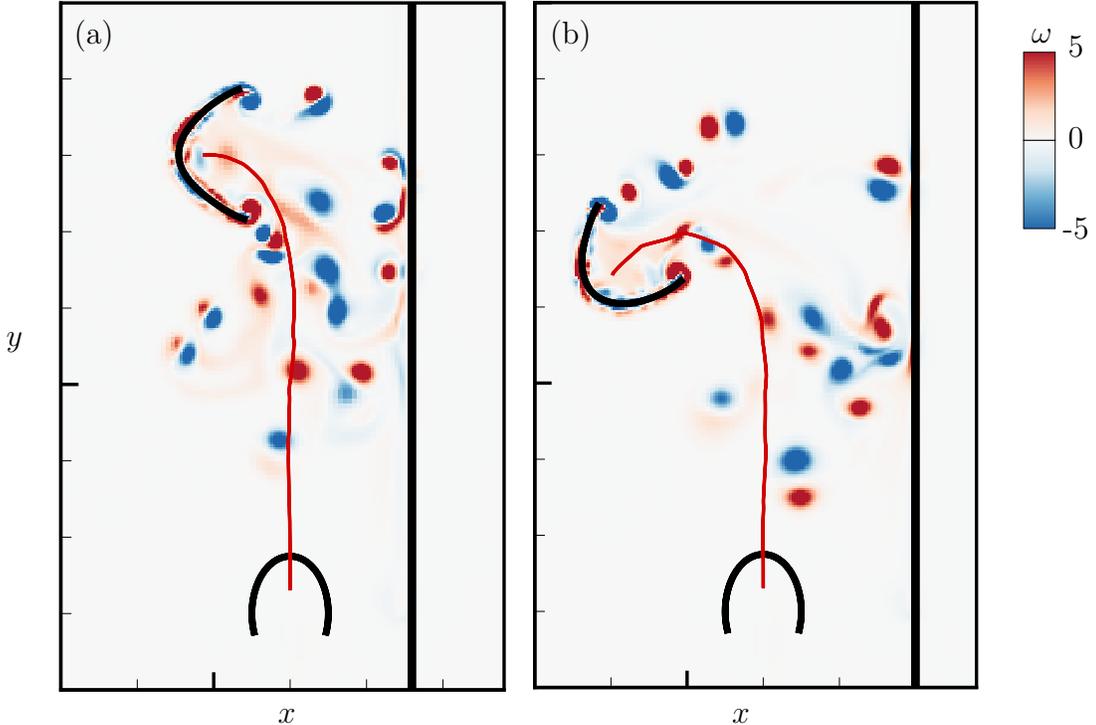
\begin{figure}
	\centering
	\begin{overpic}[scale=0.6]{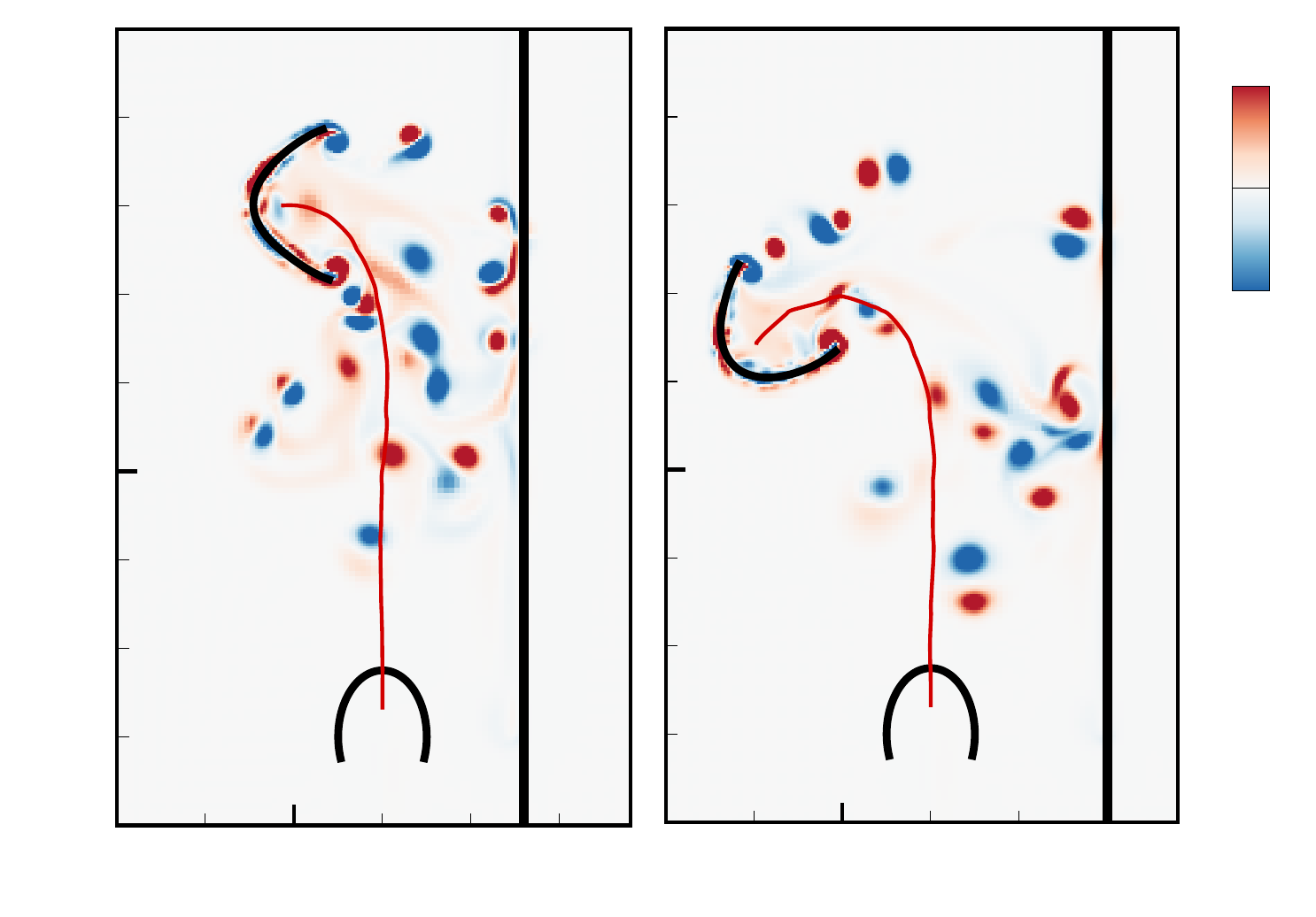}
		\put(7,4){%
			\begin{minipage}{0.4\textwidth} %
				\fontsize{12}{14}\selectfont %
				$x$ %
			\end{minipage}
		}
      \put(47,4){%
			\begin{minipage}{0.4\textwidth} %
				\fontsize{12}{14}\selectfont %
				$x$ %
			\end{minipage}
		}
      \put(-17,37){%
			\begin{minipage}{0.4\textwidth} %
				\fontsize{12}{14}\selectfont %
				$y$ %
			\end{minipage}
		}
  \put(73.5,64){%
			\begin{minipage}{0.4\textwidth} %
				\fontsize{12}{14}\selectfont %
				$\omega$ %
			\end{minipage}
		}
  \put(76.5,63){%
			\begin{minipage}{0.4\textwidth} %
				\fontsize{12}{14}\selectfont %
				5 %
			\end{minipage}
		}
  \put(76.5,55){%
			\begin{minipage}{0.4\textwidth} %
				\fontsize{12}{14}\selectfont %
				0 %
			\end{minipage}
		}
  \put(76.5,47){%
			\begin{minipage}{0.4\textwidth} %
				\fontsize{12}{14}\selectfont %
				-5  
			\end{minipage}
		}
  \put(32,64){%
			\begin{minipage}{0.4\textwidth} %
				\fontsize{12}{14}\selectfont %
				(b)  
			\end{minipage}
		}
  \put(-10,64){%
			\begin{minipage}{0.4\textwidth} %
				\fontsize{12}{14}\selectfont %
				(a)  
			\end{minipage}
		}
\end{overpic}
	\caption{Trajectory (red line) of the swimmer's forward motion over duration of $t/T=10$ at $Re=1000$, along with surrounding vorticity. (a) $d_w/d_0=1.5$. (b) $d_w/d_0=2$. The existence of the wall with different initial wall distances interacts differently with the shedding vortex from the swimmer and influences the deflection angle. Supplementary movie 1 illustrates the comparision between the two scenarios.}\label{fig:Re1000Traj} 
\end{figure}

At high $Re$, the trajectory becomes complex and unpredictable. 
In Fig.~\ref{fig:wall_perturb_traj}(b), the deflection is significantly larger and even increases at intermediate wall distance $d_w/d_0=2$ before decaying for large initial wall distances. 
Figure~\ref{fig:Re1000Traj} depicts the trajectories of the swimmer at $Re=1000$ with $d_w/d_0=1.5$ and $2.0$ (also see supplementary movie 1). 

The unpredictable trajectory of the swimmer is due to the interaction of wake vortices and the wall \citep{Terrington2024vortexWall}. 
The swimmer can generate a series vortices at moderate $Re$, or even turbulent wake at high $Re$. 
At a critical intermediate distance, the wall reflects and reorganizes the swimmer body's shed vortices.  
This interaction disrupts the stable, periodic vortex shedding pattern that would occur in an unbounded flow. 
The result is a feedback mechanism where the distorted vorticity field asymmetrically modifies the pressure distribution around the body, particularly its aft section, leading to an amplified and potentially unsteady lateral force. 
This interaction is weakened with the separation between the swimmer and the wall, as the deflection angle goes to zero at larger $d_w/d_0$. 

\begin{figure}
	\centering
	\begin{overpic}[scale=0.55]{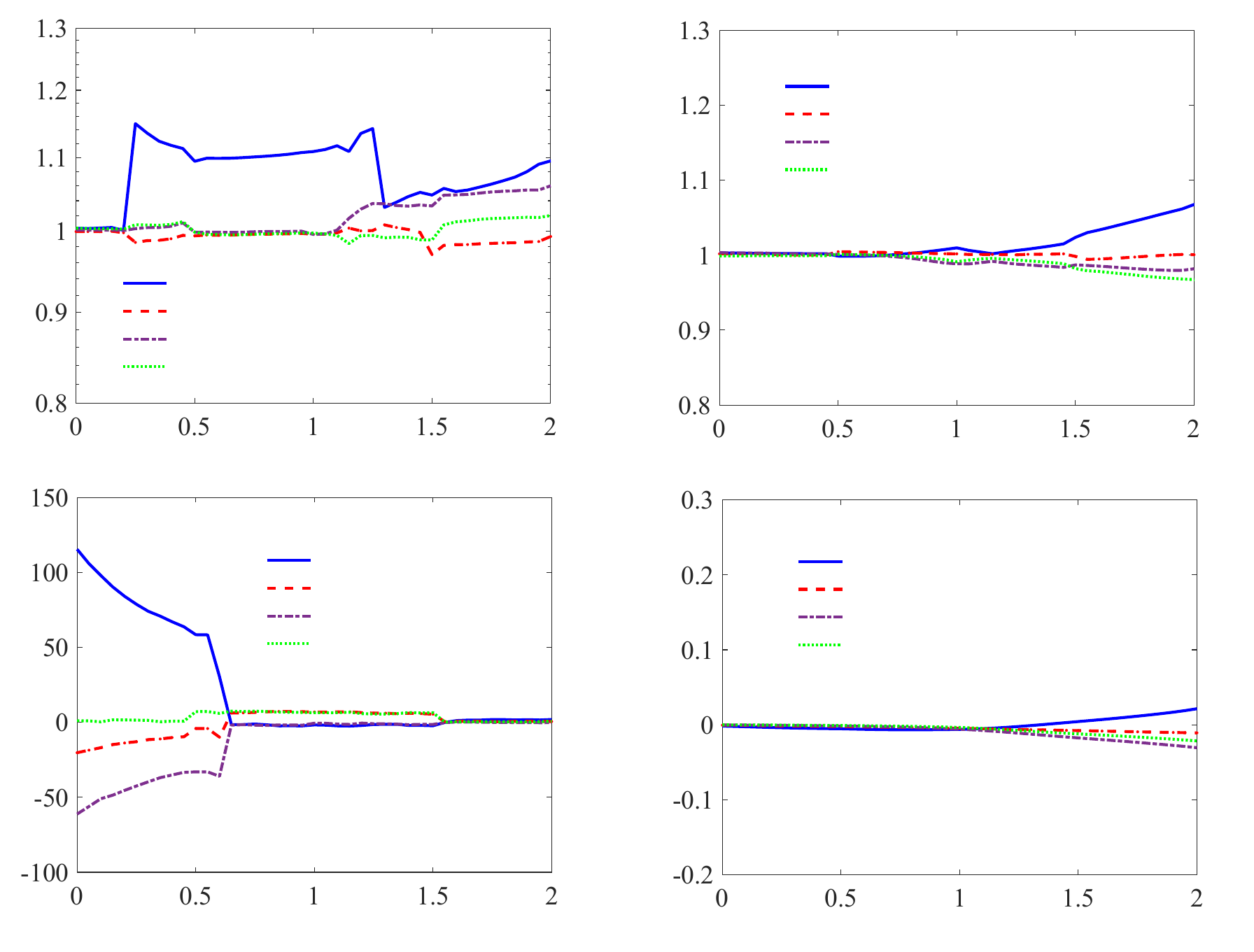}
		\put(6,0){%
			\begin{minipage}{0.4\textwidth} %
				\fontsize{12}{14}\selectfont %
				$t/T$ %
			\end{minipage}
		}
      \put(6,38){%
			\begin{minipage}{0.4\textwidth} %
				\fontsize{12}{14}\selectfont %
				$t/T$ %
			\end{minipage}
		}
      \put(57,0){%
			\begin{minipage}{0.4\textwidth} %
				\fontsize{12}{14}\selectfont %
				$t/T$ %
			\end{minipage}
		}
      \put(57,38){%
			\begin{minipage}{0.4\textwidth} %
				\fontsize{12}{14}\selectfont %
				$t/T$ %
			\end{minipage}
		}
      \put(31,23){%
			\begin{minipage}{0.4\textwidth} %
				\fontsize{12}{14}\selectfont %
				$\frac{\Delta d_w}{d_0}$ %
			\end{minipage}
		}
      \put(-21,23){%
			\begin{minipage}{0.4\textwidth} %
				\fontsize{12}{14}\selectfont %
				$\frac{M}{M_{ref}}$ %
			\end{minipage}
		}
      \put(30,59){%
			\begin{minipage}{0.4\textwidth} %
				\fontsize{12}{14}\selectfont %
				$\frac{\alpha}{\alpha_{ref}}$ %
			\end{minipage}
		}
      \put(-22,59){%
			\begin{minipage}{0.4\textwidth} %
				\fontsize{12}{14}\selectfont %
				$\frac{|\boldsymbol{F}|}{|\boldsymbol{F}_{ref}|}$ %
			\end{minipage}
		}
      \put(28,36){%
			\begin{minipage}{0.4\textwidth} %
				\fontsize{12}{14}\selectfont %
				(d) %
			\end{minipage}
		}
  \put(-22,36){%
			\begin{minipage}{0.4\textwidth} %
				\fontsize{12}{14}\selectfont %
				(c) %
			\end{minipage}
		}
      \put(28,73){%
			\begin{minipage}{0.4\textwidth} %
				\fontsize{12}{14}\selectfont %
				(b) %
			\end{minipage}
		}
  \put(-22,73){%
			\begin{minipage}{0.4\textwidth} %
				\fontsize{12}{14}\selectfont %
				(a) %
			\end{minipage}
		}
  \put(-1.5,53){%
			\begin{minipage}{0.4\textwidth} %
				\fontsize{10}{14}\selectfont %
				$d_w/d_0=1$ 
			\end{minipage}
		}
      \put(-0.5,50.75){%
			\begin{minipage}{0.4\textwidth} %
				\fontsize{10}{14}\selectfont %
				$d_w/d_0=1.5$ 
			\end{minipage}
		}
  \put(-1.5,48.5){%
			\begin{minipage}{0.4\textwidth} %
				\fontsize{10}{14}\selectfont %
				$d_w/d_0=2$ 
			\end{minipage}
		}
  \put(-0.5,46.25){%
			\begin{minipage}{0.4\textwidth} %
				\fontsize{10}{14}\selectfont %
				$d_w/d_0=2.5$ 
			\end{minipage}
		}
  \put(51.25,68.5){%
			\begin{minipage}{0.4\textwidth} %
				\fontsize{10}{14}\selectfont %
				$d_w/d_0=1$ 
			\end{minipage}
		}
  \put(52.25,66.25){%
			\begin{minipage}{0.4\textwidth} %
				\fontsize{10}{14}\selectfont %
				$d_w/d_0=1.5$ 
			\end{minipage}
		}
  \put(51.25,64){%
			\begin{minipage}{0.4\textwidth} %
				\fontsize{10}{14}\selectfont %
				$d_w/d_0=2$ 
			\end{minipage}
		}
  \put(52.25,61.75){%
			\begin{minipage}{0.4\textwidth} %
				\fontsize{10}{14}\selectfont %
				$d_w/d_0=2.5$ 
			\end{minipage}
		}
  \put(10,30.75){%
			\begin{minipage}{0.4\textwidth} %
				\fontsize{10}{14}\selectfont %
				$d_w/d_0=1$ 
			\end{minipage}
		}
  \put(11,28.5){%
			\begin{minipage}{0.4\textwidth} %
				\fontsize{10}{14}\selectfont %
				$d_w/d_0=1.5$ 
			\end{minipage}
		}
  \put(10,26.25){%
			\begin{minipage}{0.4\textwidth} %
				\fontsize{10}{14}\selectfont %
				$d_w/d_0=2$ 
			\end{minipage}
		}
  \put(11,24){%
			\begin{minipage}{0.4\textwidth} %
				\fontsize{10}{14}\selectfont %
				$d_w/d_0=2.5$ 
			\end{minipage}
		}
  \put(52.5,30.75){%
			\begin{minipage}{0.4\textwidth} %
				\fontsize{10}{14}\selectfont %
				$d_w/d_0=1$ 
			\end{minipage}
		}
  \put(53.5,28.5){%
			\begin{minipage}{0.4\textwidth} %
				\fontsize{10}{14}\selectfont %
				$d_w/d_0=1.5$ 
			\end{minipage}
		}
  \put(52.5,26.25){%
			\begin{minipage}{0.4\textwidth} %
				\fontsize{10}{14}\selectfont %
				$d_w/d_0=2$ 
			\end{minipage}
		}
  \put(53.5,24){%
			\begin{minipage}{0.4\textwidth} %
				\fontsize{10}{14}\selectfont %
				$d_w/d_0=2.5$ 
			\end{minipage}
		}
\end{overpic}
	\caption{Time evolution of (a) the swimmer's force magnitude, (b) force direction angle, and (c) torque, which are normalized by their wall-free values. (d) Corresponding displacement in the $x$-direction. All simulations are conducted at $Re=100$ with various $d_w/d_0$. 
 }\label{fig:force_angle_comp}
\end{figure}
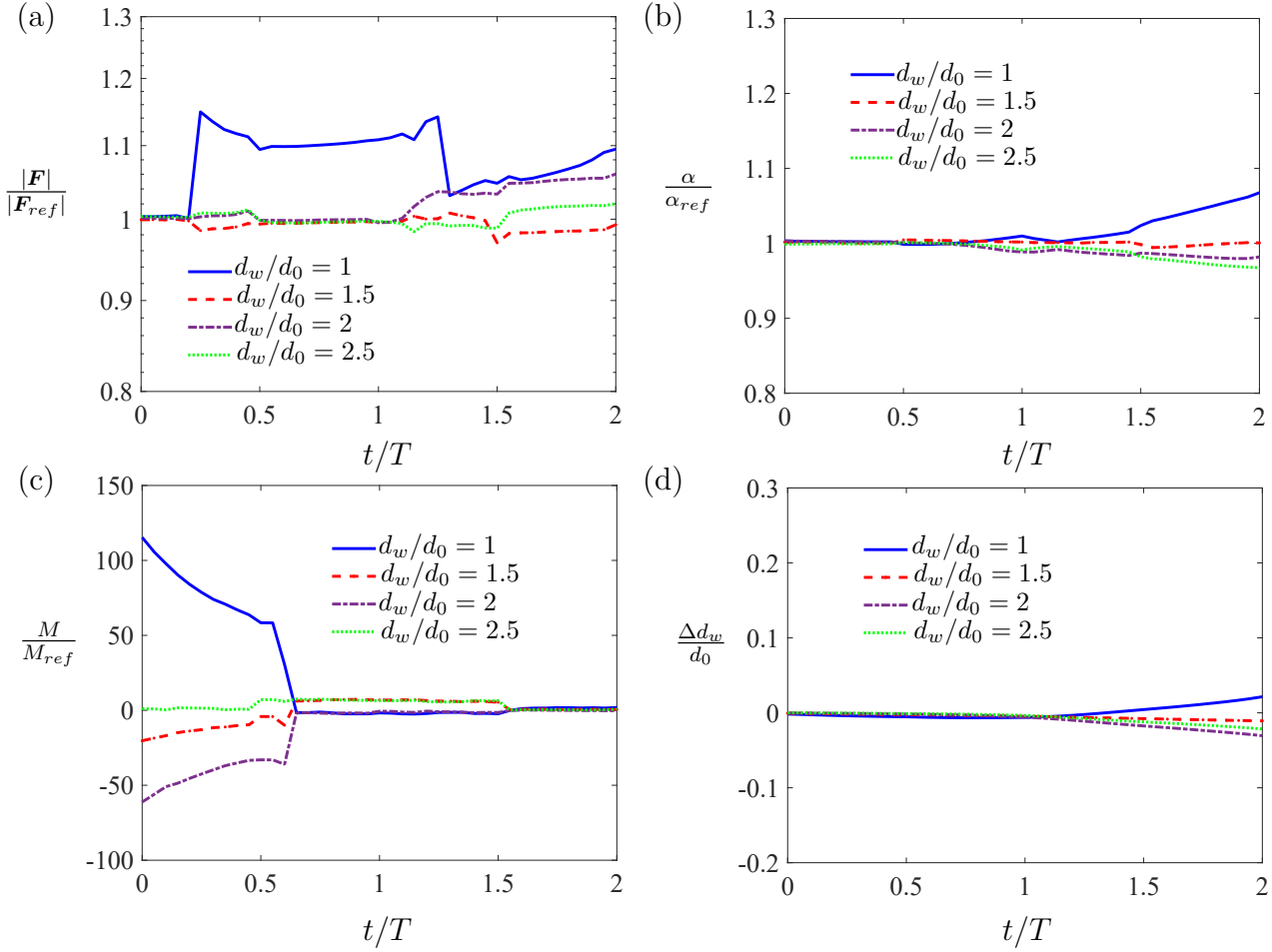

Next, we examine the differences in the swimmer’s force and torque patterns in the presence and absence of a wall. Figure~\ref{fig:force_angle_comp} compares these quantities against the wall-free case. 
Since the orientation and surrounding flow is significantly different at later times, these quantities are only comparable at early times $t/T \le 2$, where the the only difference is $d_w$. 
In Figs.~\ref{fig:force_angle_comp}(a) and (b), both the magnitude and direction of the force for different $d_r$ deviate at $t/T \le 2$.
Figure~\ref{fig:force_angle_comp}(c) reveals that the torque differs substantially even at very early times $t/T \le 0.5$, while Fig.~\ref{fig:force_angle_comp}(d) indicates that the displacement in the 
$x$-direction remains nearly zero. 
Despite the distance $d_w/d_0\ge 2$, a small but non-zero negative x-displacement is observed, highlighting the complex vortex-wall interactions. 
These distinctions in force and torque demonstrate the influence of the vortex-wall interaction when the swimmer is near a wall.

\begin{figure}
	\centering
	\begin{overpic}[scale=0.6]{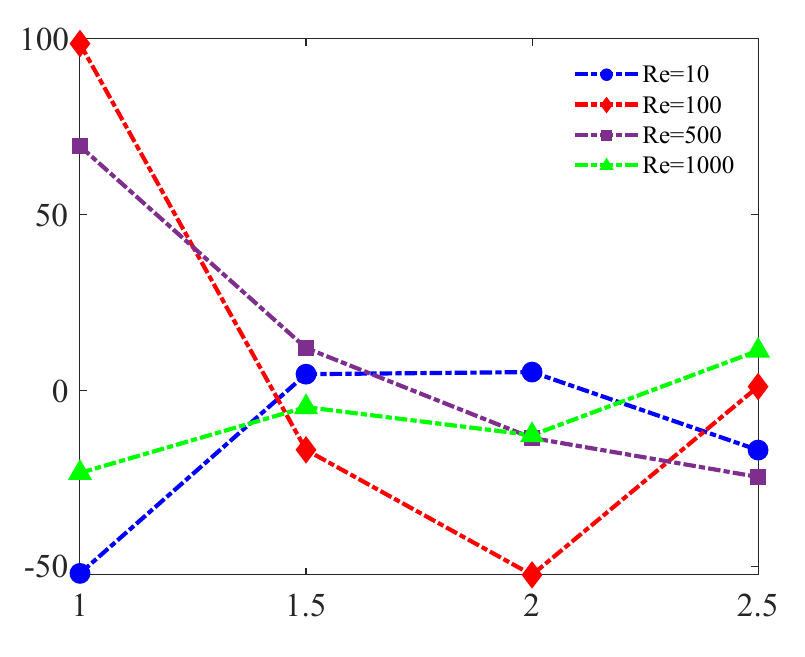}
		\put(14,-2){%
			\begin{minipage}{0.4\textwidth} %
				\fontsize{12}{14}\selectfont %
				$d_w/d_0$ %
			\end{minipage}
		}
      \put(-39,38){%
			\begin{minipage}{0.4\textwidth} %
				\fontsize{12}{14}\selectfont %
				$\frac{M}{M_{ref}}$ %
			\end{minipage}
		}
\end{overpic}
	\caption{The swimmer's torque at beginning stage normalized by the wall-free case for different $d_w$ and $Re$. 
 }\label{fig:torque_plot}
\end{figure}

In particular, Fig.~\ref{fig:force_angle_comp}(c) indicates that the notable torque acting on the swimmer at $Re=100$ deviates significantly from nearly zero torque in the wall-free scenario at the initial stage, regardless of the specific wall distance. 
Furthermore, Fig.~\ref{fig:torque_plot} plots the initial torque at $t/T=0.1$ as a function of $d_w/d_0$ for various $Re$, confirming that wall proximity induces substantial and non-monotonic changes in torque. 
The strong dependence on both $d_w$ and $Re$ highlights the significance of torque. Since the torque determines the moving direction of the swimmer, it is essential to be perceived by an agent with measuring its hydrodynamic surroundings accurately.
Additionally, we examined the effect of wall when it is put in front of the swimmer, and found that the presence of a front wall has minor influence on the propulsion of the swimmer (not shown).

\subsection{\label{sec:obs}Wall effects in obstacle avoidance}
\begin{figure}
	\centering
	\begin{overpic}[scale=0.45]{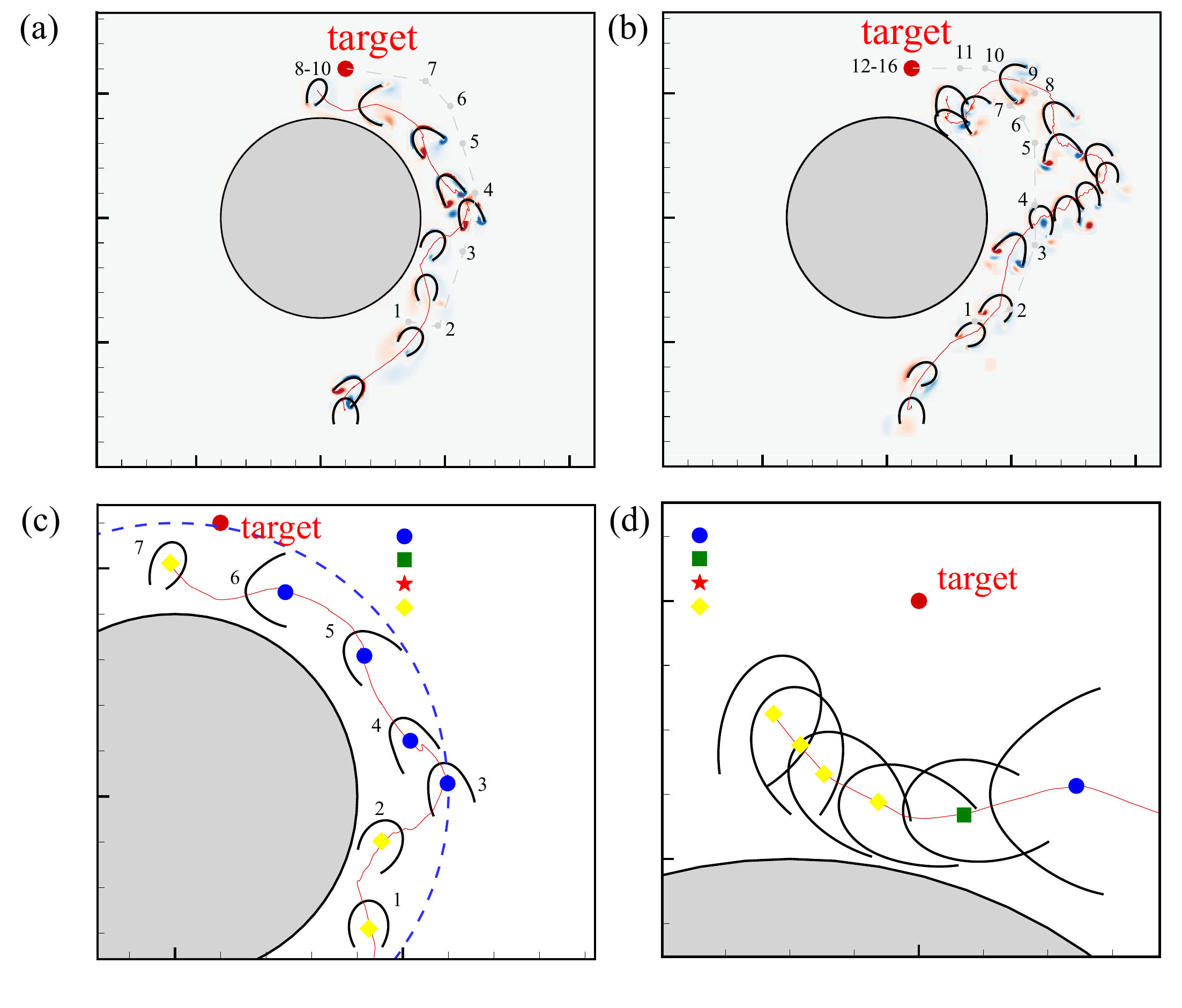}
		\put(11,1){%
			\begin{minipage}{0.4\textwidth} %
				\fontsize{14}{14}\selectfont %
				$x$ %
			\end{minipage}
		}
      \put(57,1){%
			\begin{minipage}{0.4\textwidth} %
				\fontsize{14}{14}\selectfont %
				$x$ %
			\end{minipage}
		}
      \put(-14,22){%
			\begin{minipage}{0.4\textwidth} %
				\fontsize{14}{14}\selectfont %
				$y$ %
			\end{minipage}
		}
      \put(-14,63){%
			\begin{minipage}{0.4\textwidth} %
				\fontsize{14}{14}\selectfont %
				$y$ %
			\end{minipage}
		}
      \put(22.8,39.2){%
			\begin{minipage}{0.4\textwidth} %
				\fontsize{9}{14}\selectfont %
				$A_0$(move forward)%
			\end{minipage}
		}
      \put(21.3,37.2){%
			\begin{minipage}{0.4\textwidth} %
				\fontsize{9}{14}\selectfont %
				$A_1$(turn right)%
			\end{minipage}
		}
      \put(20.7,35.2){%
			\begin{minipage}{0.4\textwidth} %
				\fontsize{9}{14}\selectfont %
				$A_2$(turn left)%
			\end{minipage}
		}
      \put(19,33.2){%
			\begin{minipage}{0.4\textwidth} %
				\fontsize{9}{14}\selectfont %
				$A_3$(drift)%
			\end{minipage}
		}
      \put(47.8,39.2){%
			\begin{minipage}{0.4\textwidth} %
				\fontsize{9}{14}\selectfont %
				$A_0$(move forward)%
			\end{minipage}
		}
      \put(46.3,37.2){%
			\begin{minipage}{0.4\textwidth} %
				\fontsize{9}{14}\selectfont %
				$A_1$(turn right)%
			\end{minipage}
		}
      \put(45.7,35.2){%
			\begin{minipage}{0.4\textwidth} %
				\fontsize{9}{14}\selectfont %
				$A_2$(turn left)%
			\end{minipage}
		}
      \put(44,33.2){%
			\begin{minipage}{0.4\textwidth} %
				\fontsize{9}{14}\selectfont %
				$A_3$(drift)%
			\end{minipage}
		}
      \put(23,28.3){%
			\begin{tikzpicture}[overlay,scale=0.52]
                \draw[line width=1pt, color=red][->] (-2.55,3)--(-2.15,3.6);
                \draw[line width=1pt, color=red][->] (0.55,2.15)--(1.15,2.35);
                \draw[line width=1pt, color=red][->] (2.3,0.35)--(0.9,1.75);
                \draw[line width=1pt, color=red][->] (3.6,-1.9)--(2.4,-0.5);
                \draw[line width=1pt, color=red][->] (4.6,-3.2)--(3.35,-1.8);
                \draw[line width=1pt, color=red][->] (2.75,-4.55)--(1.75,-5.35);
                \draw[line width=1pt, color=red][->] (2.9,-6.9)--(3.65,-7.6);
			\end{tikzpicture}
		}
      \put(70.5,29){%
			\begin{tikzpicture}[overlay,scale=0.53]
                \draw[line width=1pt, color=red][->] (-1.5,-1.2)--(-1.4695,-0.915);
                \draw[line width=1pt, color=red][->] (-0.85,-2.15)--(-0.63,-2.4);
                \draw[line width=1pt, color=red][->] (-0.2,-2.75)--(-0.1,-2.5);
                \draw[line width=1pt, color=red][->] (1.1,-3.4)--(0.5,-2.6);
                \draw[line width=1pt, color=red][->] (3.55,-3.8)--(3.7,-3.45);
                \draw[line width=1pt, color=red][->] (6.45,-3.0)--(6.7,-2.6);
			\end{tikzpicture}
		}
\end{overpic}
	\caption{Trajectory (red line) of the swimmer and surrounding vorticity field in a simple obstacle avoidance task with (a) current SAC agent at $t/T=0,5,...,45$, and (b)previous agent \citep{chen2024DRL} without force and torque input at $t/T=0,5,...,75$. Grey dots and line with stations are pilots and their trajectory used to guide the swimmer. (c) Trajectory of the swimmer (red line) with current SAC agent and the force experienced (red arrow) and action choice over $t/T=15,20,...,45$ and (d) $t/T=40,41,...,45$. Blue dotted line represents the influence range of the obstacle.}\label{fig:BigObs}
\end{figure}

Based on our existing tracking strategy~\citep{chen2024DRL}, we incorporate a pilot which changes every time step for the swimmer to track the actual specified target.
To reach the target point and avoid obstacles, a pathfinding algorithm ``A*'' \citep{Peter1968Astar} is used to generate pilots, which is detailed in Section~\ref{sec:obstacle}. 
Figure~\ref{fig:BigObs}(a) shows the trajectory of the swimmer in a simple obstacle avoidance tracking task. 
The swimmer manages to move along the edge of the cylinder obstacle and reach the target.

As discussed for Fig.~\ref{fig:wall_perturb_traj}, the influence range of wall effect for current setup is around $2d_0$. 
Figure~\ref{fig:BigObs}(c) shows that during nearly the whole process, the swimmer is within the influence range. 
Initially, the swimmer approaches the wall and experiences a repulsive force that pushes it backward (station 1). 
It then nearly makes contact with the wall and utilizes this repulsion to facilitate a rapid turn (station 2). 
Following the turn, the swimmer consistently selects action $A_0$ to propel itself forward. 
Throughout this phase, it counteracts the repulsive force from the wall to maintain a stable distance (stations 3-6). 
As it approaches the target, the swimmer selects action $A_3$ constantly to once again leverage wall interactions, executing a final turn to achieve the target position (stations 6 and 7).

Figure~\ref{fig:BigObs}(d) details the swimmer utilizing the wall to execute a turn. 
Throughout this near-wall maneuver, action $A_3$ is consistently selected after the selection of $A_1$, allowing the swimmer to drift passively with the surrounding flow.
The turn is accelerated by the vortex-wall interaction, where flow reflections impart momentum to propel the swimmer away from the boundary. 
Combined with the velocity produced by the action $A_1$ previously selected, this enables a rapid reorientation within a confined space.

For comparison, Fig.~\ref{fig:BigObs}(b) shows the trajectory of a swimmer without force and torque sensing as agent inputs, performing the same task as in Fig.~\ref{fig:BigObs}(a). 
Without sensing the surrounding flow's force $(F_x, F_y)$ and torque $M$, this swimmer is pushed away from the obstacle by the reflected flow and fails to maintain proximity to the wall. 
It only reorients towards the target once sufficiently distant from the obstacle. 

This phenomenon can be attributed to the reinforcement learning training process. 
Consider an agent whose state is represented by a 12-dimensional vector. 
In an obstacle-free setting, when the agent is in state $\boldsymbol{s}$ with no external force $(F_x,F_y)$ or torque $M$, taking action $A$ transitions it to a new state $\boldsymbol{s}_1$ and yields an immediate reward $r$. 
The deep Q-network (DQN) algorithm adopted in Ref.~\citep{chen2024DRL} trains the agent’s Q-network to satisfy the Bellman optimality equation:
\begin{equation}
Q(\boldsymbol{s},A)=r+\max_{A^\prime} Q(\boldsymbol{s}_1,A^\prime).
\end{equation}
This equation indicates that the optimal Q-value $Q(\boldsymbol{s},A)$, the maximum cumulative reward from taking action $A$ in state $\boldsymbol{s}$, comprises two components: the immediate reward $r$, and the maximum cumulative reward achievable from the next state $\boldsymbol{s}_1$ over all possible subsequent actions $A^\prime$.
However, when obstacles are present, executing the same action $A$ in state $\boldsymbol{s}$ leads the agent to a different state $\boldsymbol{s}_2$, rather than $\boldsymbol{s}_1$. 
As a result, the Bellman optimality equation, which assumes a transition to $\boldsymbol{s}_1$, no longer holds. 
This discrepancy leads the agent to take suboptimal actions in obstacle-related states.

\begin{figure}
	\centering
	\begin{overpic}[scale=0.6]{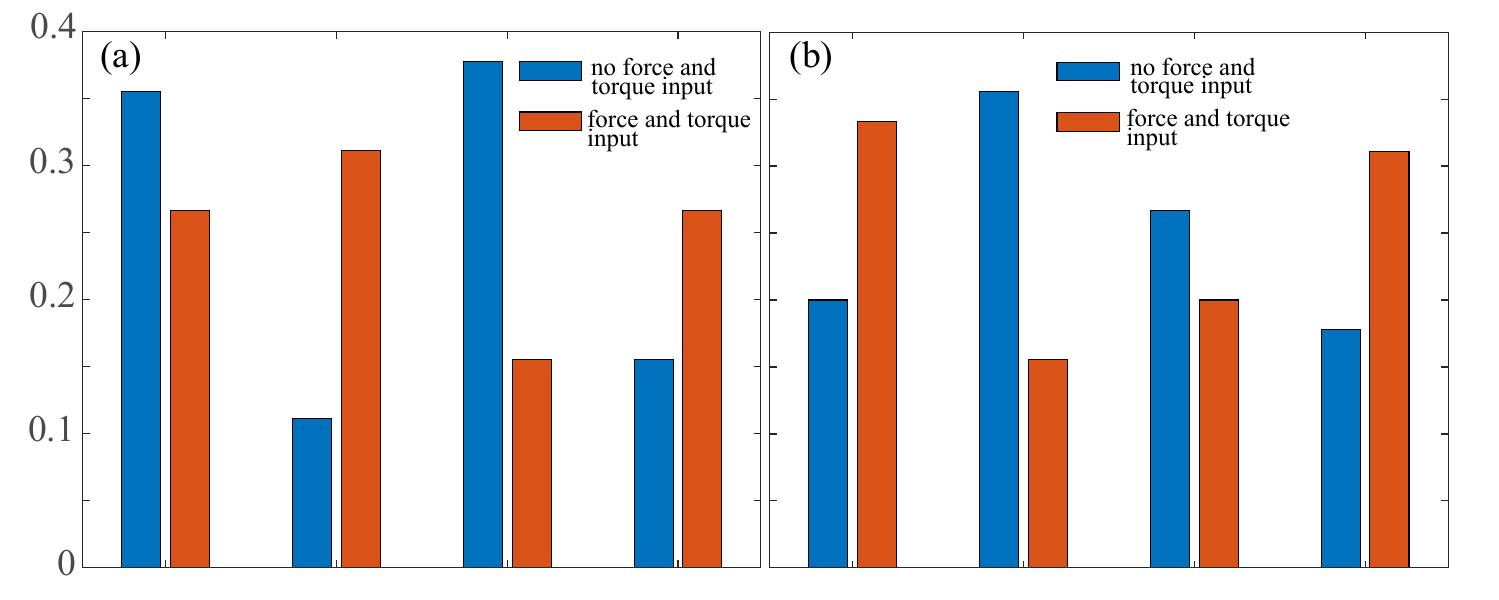}
		\put(-10.5,0){%
			\begin{minipage}{0.4\textwidth} %
				\fontsize{9}{14}\selectfont %
				$A_0$ %
			\end{minipage}
		}
      \put(1,0){%
			\begin{minipage}{0.4\textwidth} %
				\fontsize{9}{14}\selectfont %
				$A_1$ %
			\end{minipage}
		}
      \put(12.5,0){%
			\begin{minipage}{0.4\textwidth} %
				\fontsize{9}{14}\selectfont %
				$A_2$ %
			\end{minipage}
		}
      \put(24,0){%
			\begin{minipage}{0.4\textwidth} %
				\fontsize{9}{14}\selectfont %
				$A_3$ %
			\end{minipage}
		}
  \put(36,0){%
			\begin{minipage}{0.4\textwidth} %
				\fontsize{9}{14}\selectfont %
				$A_0$ %
			\end{minipage}
		}
      \put(47.5,0){%
			\begin{minipage}{0.4\textwidth} %
				\fontsize{9}{14}\selectfont %
				$A_1$ %
			\end{minipage}
		}
      \put(59,0){%
			\begin{minipage}{0.4\textwidth} %
				\fontsize{9}{14}\selectfont %
				$A_2$ %
			\end{minipage}
		}
      \put(70.5,0){%
			\begin{minipage}{0.4\textwidth} %
				\fontsize{9}{14}\selectfont %
				$A_3$ %
			\end{minipage}
		}
\end{overpic}
	\caption{Comparison of action choices histogram between models with/without force and torque input at (a) $t/T=0-22.5$ and at (b) $t/T=22.5-45$. The model with force and torque input selects more $A_1$ to turns more at the early stage and $A_0$ to stabilize its orientation.}\label{fig:action_compare}
\end{figure}
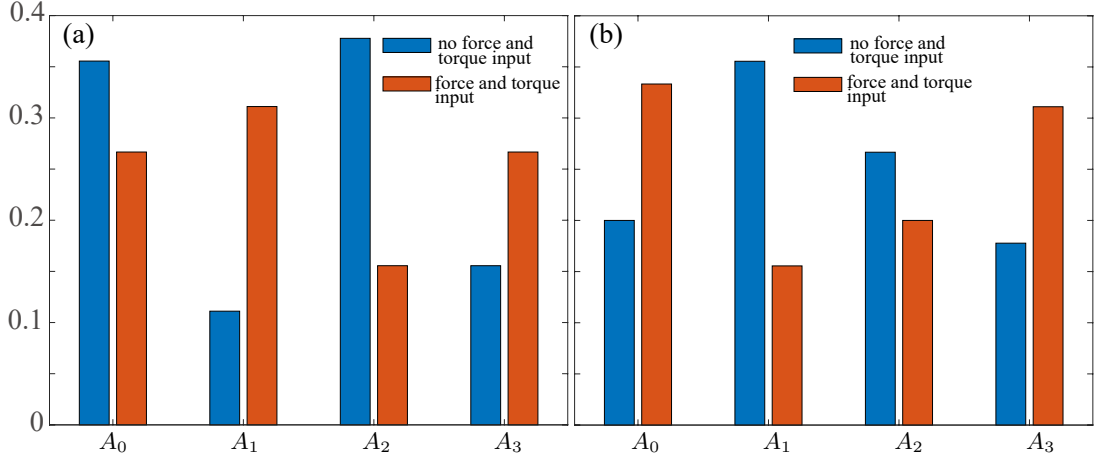

Figure~\ref{fig:action_compare} compares action selections over time. During the initial $45T$ interval, the present model (SAC agent with force and torque input) predominantly selects $A_1$, inducing sustained rightward turning that positions the swimmer distally from the obstacle. Subsequently, it reorients toward the obstacle, ultimately harnessing wall contact to rapidly shift direction. In the latter $45T$ phase, increased $A_0$ selection maintains obstacle clearance while mitigating repulsive wall-interaction forces. The maneuver concludes by exploiting wall-induced flow to align with the target direction. 
Conversely, the old model (previous agent without force and torque input) \citep{chen2024DRL} selects $A_1$ less frequently, favoring $A_2$ to pursue the target. During the second $22.5T$, wall-induced flow propels it away from the boundary, necessitating extended realignment to achieve target orientation. 

\section{\label{sec:obstacle}Navigation in environment with obstacles}
Next, we will illustrate the performance of DRL agent with force and torque input using tracking tasks in environments with complicated obstacles. 
The schematic of how the swimmer detects environment and derives pilot is shown in Fig.~\ref{fig:navi_process} (also see supplementary movie 2). 
The environment is set up within the 2D computational domain, which is discretized with a grid spacing of $\Delta x=d_0 /2$, and is effectively represented as a grid-based maze. 
In the immersed boundary method, solid walls are represented by closely spaced Lagrangian points. 
A grid cell is designated as an ``obstacle'' if it contains at least one Lagrangian point representing the obstacle boundary. 
Crucially, the swimmer possesses no prior knowledge of obstacle locations within the domain. 
Instead, it can detect obstacle Lagrangian points located within a specified sensing radius. 
To mimic realistic perception, obstacle points within this sensing range remain undetectable if they are visually blocked by other obstacles from the swimmer's viewpoint. 

\begin{figure}
	\centering
	\begin{overpic}[scale=0.5]{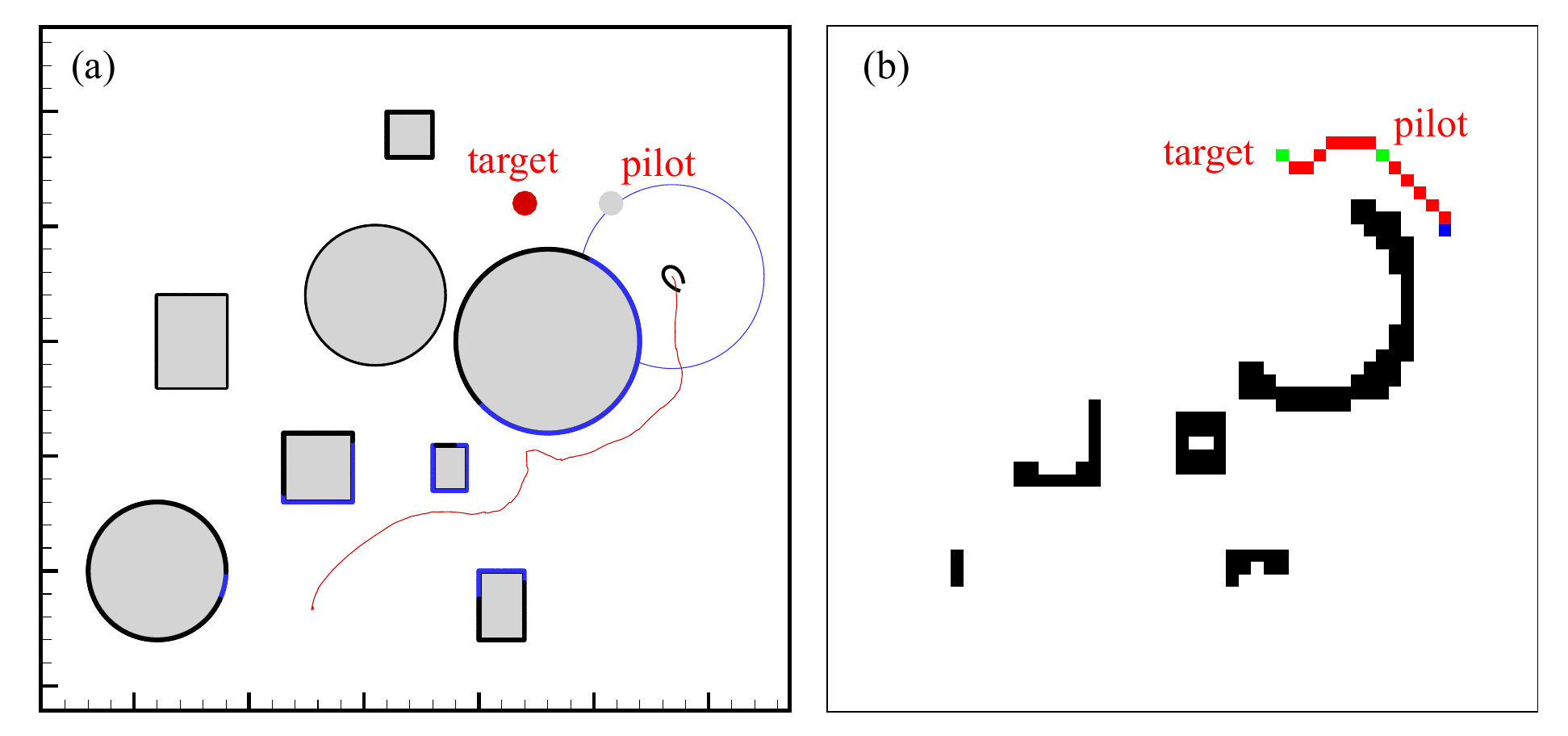}
		\put(4,-2){%
			\begin{minipage}{0.4\textwidth} %
				\fontsize{12}{14}\selectfont %
				$x$ %
			\end{minipage}
		}
      \put(55,-2){%
			\begin{minipage}{0.4\textwidth} %
				\fontsize{12}{14}\selectfont %
				$x$ %
			\end{minipage}
		}
      \put(-20.5,22){%
			\begin{minipage}{0.4\textwidth} %
				\fontsize{12}{14}\selectfont %
				$y$ %
			\end{minipage}
		}
\end{overpic}
	\caption{(a) Trajectory of the swimmer (red line). The swimmer has a sensing range of $\mathcal{R}=4d_0$ (blue circle). The boundaries sensed by the swimmer are updated in blue. (b) Based on the sensed boundaries (black dots), the domain is descretized, and A* algorithm is applied to generate path (red dots) from the swimmer (blue dot) to the target with a pilot (green dots) . The pilot is chosen such that it is on the path and has an approximate distance of $\mathcal{R}$ from the swimmer. The tracking process is presented in supplementary movie 2.}\label{fig:navi_process} 
\end{figure}

Given this grid maze representation with locally sensed obstacle information, the A* algorithm (detailed in Appendix~\ref{sec:Astar}) is employed for pathfinding. 
This algorithm generates an optimal path from the swimmer's current position to a designated target. 
Subsequently, a path-following approach is utilized to derive a pilot for the swimmer to track along this computed path.
Note that the map detection and swimming are conducted together and the swimmer gradually detect its surroundings.

We will first assess the swimmer's locomotion and obstacle avoidance capabilities in an open flow domain containing obstacles (external flow scenario). 
Subsequently, to evaluate performance for real-world applications such as cave diving exploration, the swimmer navigates towards a predefined target with navigating complex, enclosed, maze-like domains (internal flow scenario). 

\subsection{\label{sec:outflow}Navigation in external flow}
\begin{figure}
	\centering
	\begin{overpic}[scale=0.6]{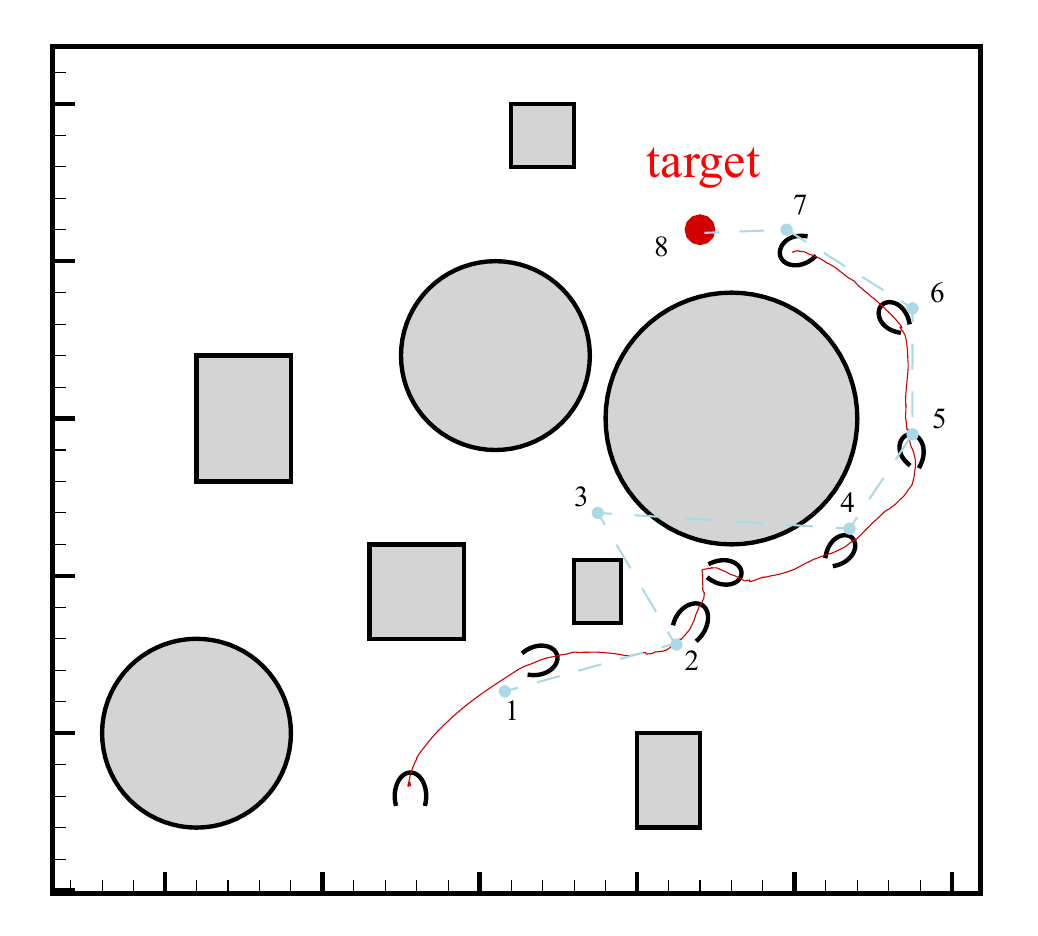}
		\put(20,-2){%
			\begin{minipage}{0.4\textwidth} %
				\fontsize{14}{14}\selectfont %
				$x$ %
			\end{minipage}
		}
      \put(-33,42){%
			\begin{minipage}{0.4\textwidth} %
				\fontsize{14}{14}\selectfont %
				$y$ %
			\end{minipage}
		}
\end{overpic}
	\caption{Trajectory (red line) of swimmer at $t/T$=0,9,...,63. Light blue dots and line with stations are pilots and their trajectory used to guide the swimmer, respectively.}\label{fig:outter_traj} 
\end{figure}

The swimmer navigates an open environment containing unknown, randomly distributed, and irregular obstacles. 
Figure~\ref{fig:outter_traj} depicts its trajectory.
It initially advance northeast until encountering the first obstacle, where hydrodynamic repulsion forces induce backward displacement. 
Subsequently, it circumvents the obstacle and resumes forward propulsion. At $t/T=27$, collision with a second obstacle occurs, and the swimmer harnesses this interaction to rapidly execute a rightward turn via wall-induced vorticity. 
It then maintains boundary-proximal locomotion until successfully reaching the target.

\begin{figure}
	\centering
	\begin{overpic}[scale=0.4]{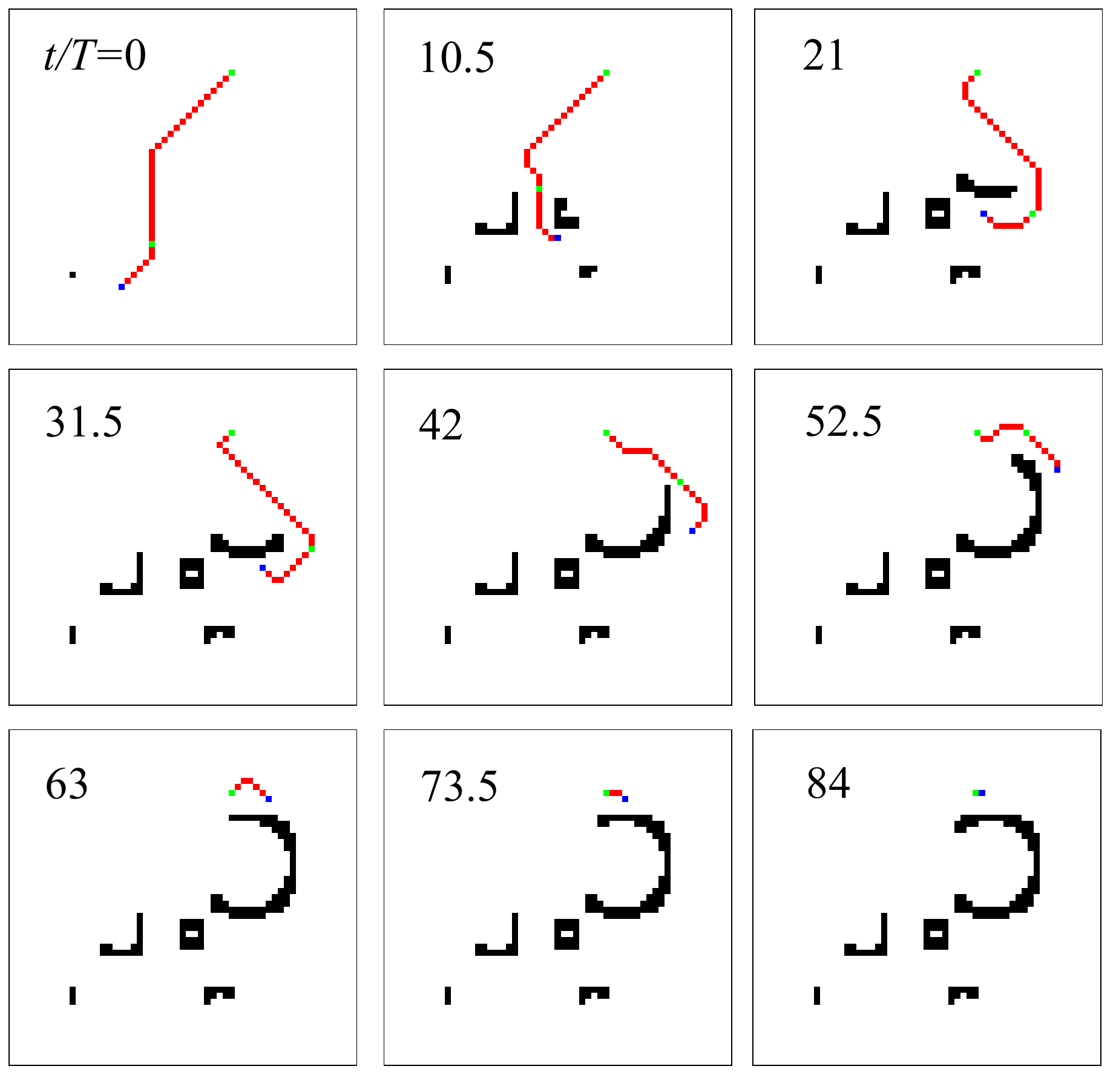}
		\put(24,-3){%
			\begin{minipage}{0.4\textwidth} %
				\fontsize{14}{14}\selectfont %
				$x$ %
			\end{minipage}
		}
      \put(-30,46){%
			\begin{minipage}{0.4\textwidth} %
				\fontsize{14}{14}\selectfont %
				$y$ %
			\end{minipage}
		}
      \put(62,83){%
			\begin{tikzpicture}[overlay,scale=0.53]
                \draw[line width=1pt, color=red][->] (-1.2,0.9)--(-2.2,0.8);
                \draw[line width=1pt, color=red][->] (4.6,0.9)--(5.6,0.8);
			\end{tikzpicture}
		}
      \put(43,86){%
			\begin{minipage}{0.4\textwidth} %
				\fontsize{12}{14}\selectfont %
				path replanning  
			\end{minipage}
		}
\end{overpic}
	\caption{Maze map constructed in real time with position of the swimmer (blue dot), planned path (red dots), pilot and target (both in green dots) at different $t/T$ (marked at the left upper of each panel)}.\label{fig:outflow_Astar} 
\end{figure}
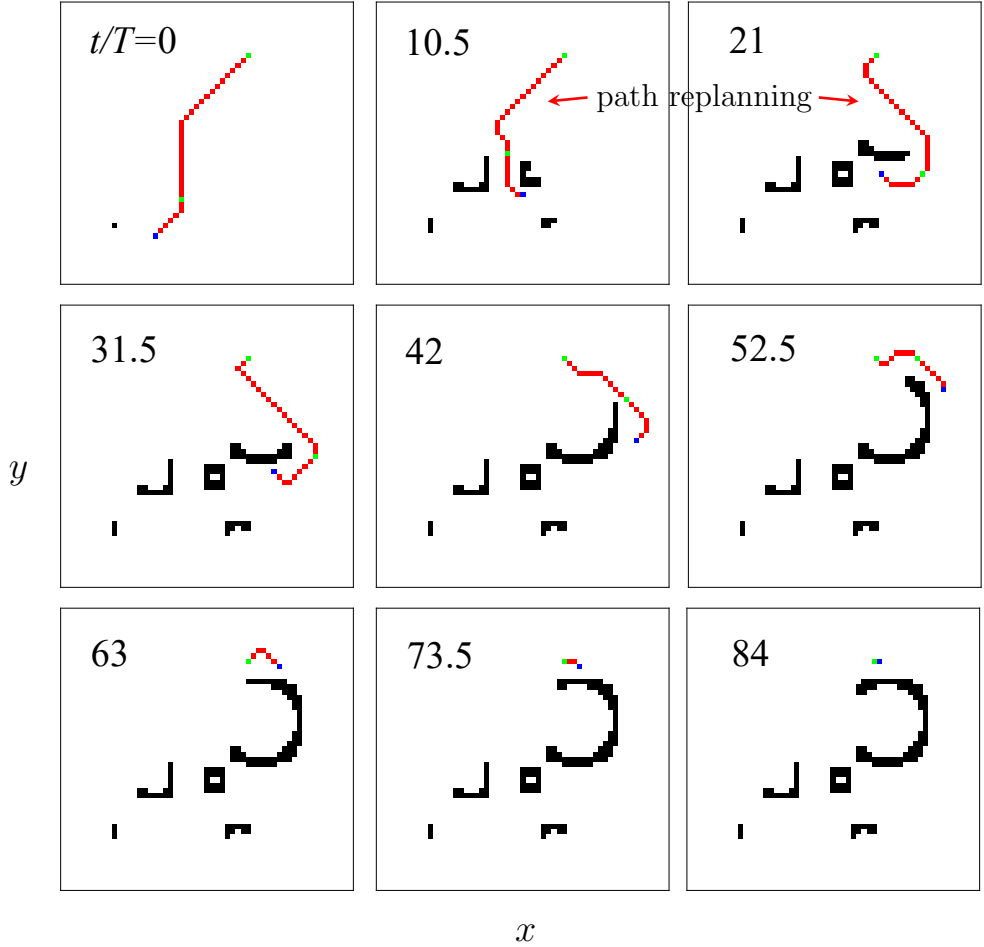

The path planning sequence is illustrated in Fig.~\ref{fig:outflow_Astar}. 
Initially, the A* algorithm generates a path through the gap between two detected obstacles. 
However, due to inertia, the swimmer deviates to the right side of these obstacles during $t/T=9-18$, triggering the first path replanning. 
As shown by the pilot at station 3 of Fig.~\ref{fig:outter_traj}, the path is initially adjusted to pass between the square and cylindrical obstacles. 
A second replanning is then activated, shifting the path further rightward. 
This newly computed path proves more effective by better balancing the dual objectives of trajectory length minimization and obstacle avoidance.
Following the replanning, the pilot proceeds along the cylindrical obstacle and successfully guides the swimmer toward the actual target.

\subsection{\label{sec:simple_cave_given_target}Exploration in internal flow}
In real-world underwater exploration or rescue missions, targets are often unknown in advance.
Therefore, the swimmer must conduct a comprehensive exploration of the environment to detect potential targets of interest.
We model the geometry of a real cave \citep{Povara2019} and define the task as fully exploring the cave and returning to the starting position.
The swimmer has a sensing range of $\mathcal{R} = 4d_0$, which is similar to the box jellyfish Tripedalia cystophora in natural environment \citep{BIELECKI2023boxJelly}.
Boundaries within this range that are not obstructed by other boundaries are considered explored. 
The cave modeled with a closed boundary is considered fully explored once its entire boundary has been covered.

Figure~\ref{fig:simple_traj} illustrates the trajectory of the swimmer during the exploration process (also see supplementary movie 3).
The swimmer initially advances forward to conduct exploration of the environment. When it encounters a fork in the path in Fig.~\ref{fig:simple_traj}(a), it opts to proceed along the left branch first. 
After reaching the end of this branch, the swimmer backtracks and then investigates the alternative fork. 
Once this section has been thoroughly explored, the entire cave is fully mapped, prompting the swimmer to begin its return journey to the starting point.
During the return navigation through the narrow passage leading to the origin, the swimmer's trajectory becomes noticeably sinuous due to the wall effect. 
This effect induces complex hydrodynamic interactions, causing deviations in the intended path. 
Despite being pushed backward on two occasions by the surrounding flow, the swimmer demonstrates robust navigational capability and succeeds in returning to the origin.

\begin{figure}
	\centering
	\begin{overpic}[scale=0.47]{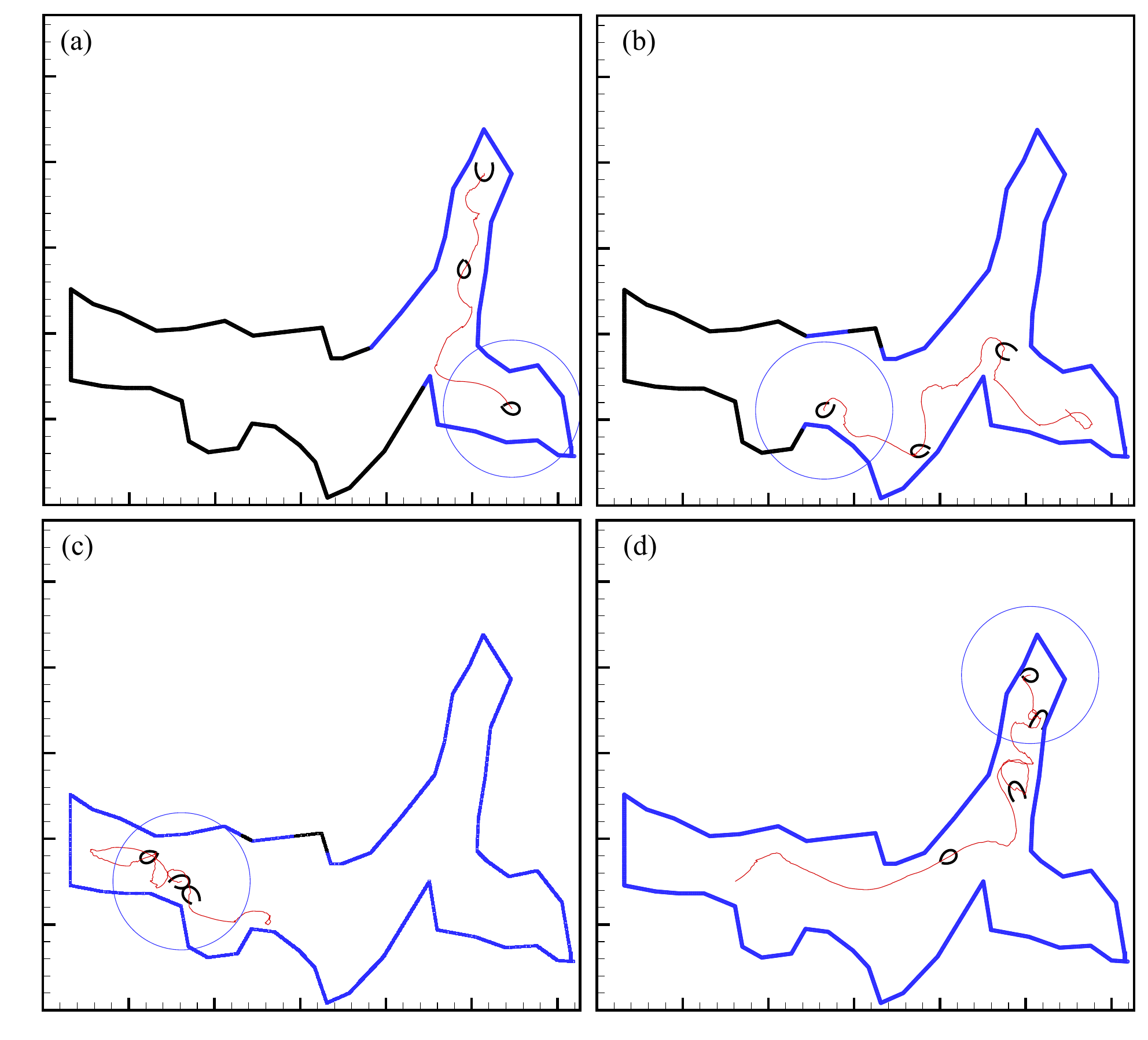}
		\put(7,0){%
			\begin{minipage}{0.4\textwidth} %
				\fontsize{12}{14}\selectfont %
				$x$ %
			\end{minipage}
		}
      \put(58,0){%
			\begin{minipage}{0.4\textwidth} %
				\fontsize{12}{14}\selectfont %
				$x$ %
			\end{minipage}
		}
      \put(-19,22){%
			\begin{minipage}{0.4\textwidth} %
				\fontsize{12}{14}\selectfont %
				$y$ %
			\end{minipage}
		}
      \put(-19,67){%
			\begin{minipage}{0.4\textwidth} %
				\fontsize{12}{14}\selectfont %
				$y$ %
			\end{minipage}
		}
\end{overpic}
	\caption{Trajectory of swimmer (in red line) at (a) $t/T=0,25,50$. (b) $t/T=75,100,125$. (c) $t/T=150,175,200$. (d) $t/T=225,250,275,300$. The sensing range is represented by the blue circle. Explored boundary is updated in blue. Supplementary movie 3 illustrates the exploring process.}\label{fig:simple_traj}
\end{figure}
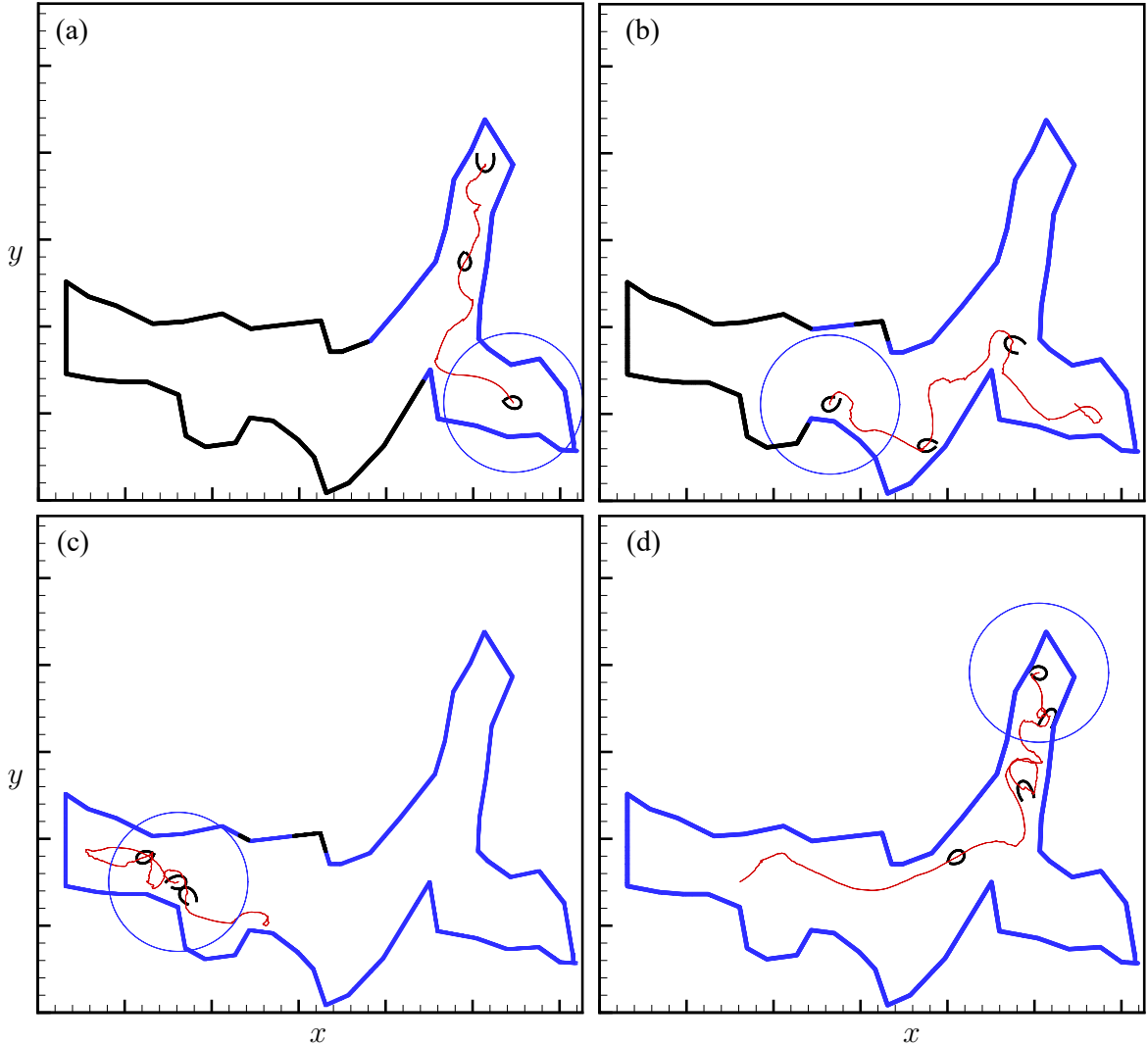

\begin{figure}
	\centering
	\begin{overpic}[scale=0.45]{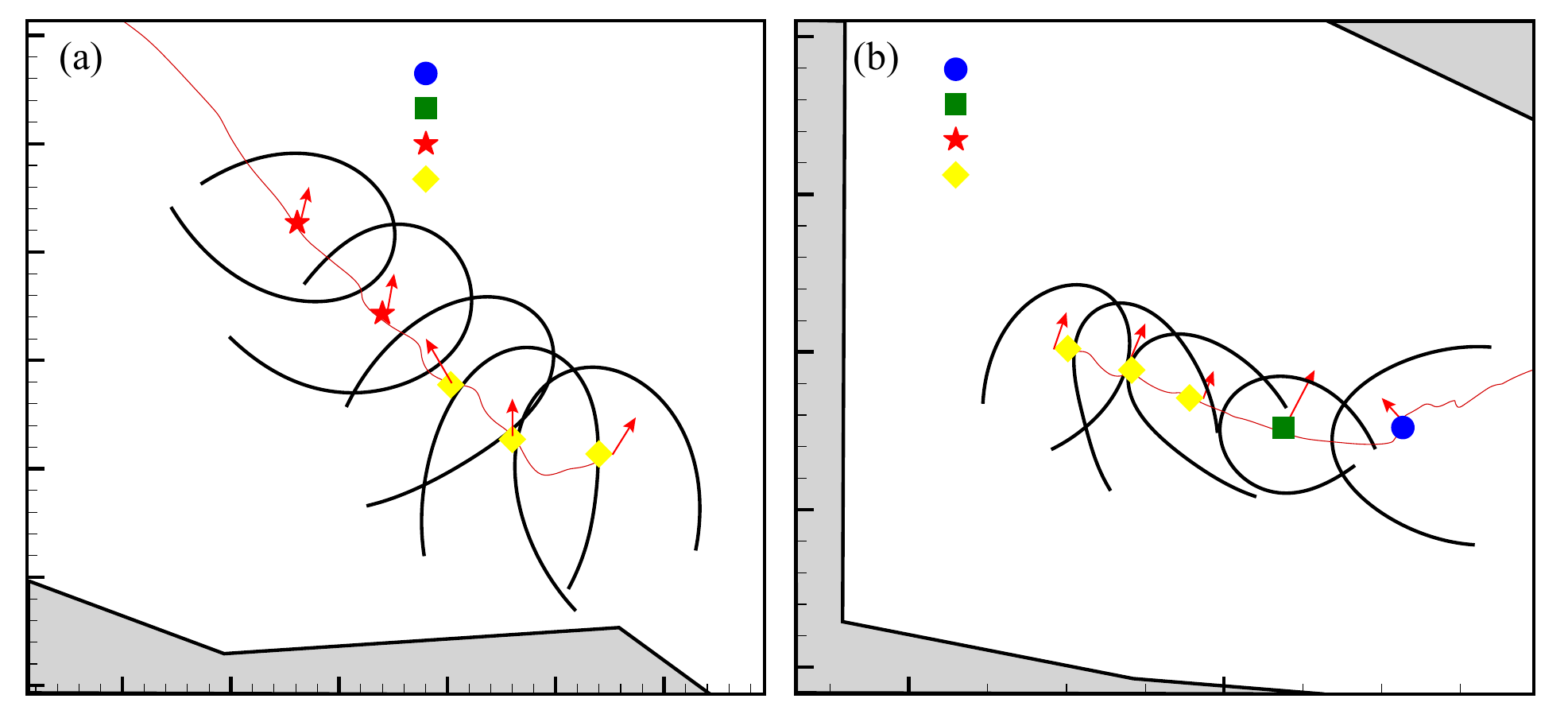}
		\put(6,-1){%
			\begin{minipage}{0.4\textwidth} %
				\fontsize{12}{14}\selectfont %
				$x$ %
			\end{minipage}
		}
      \put(54,-1){%
			\begin{minipage}{0.4\textwidth} %
				\fontsize{12}{14}\selectfont %
				$x$ %
			\end{minipage}
		}
      \put(-22,20){%
			\begin{minipage}{0.4\textwidth} %
				\fontsize{12}{14}\selectfont %
				$y$ %
			\end{minipage}
		}

      \put(14.5,40.5){
			\begin{minipage}{0.4\textwidth} %
				\fontsize{9}{14}\selectfont %
				$A_0$(move forward)%
			\end{minipage}
		}
      \put(13,38.2){%
			\begin{minipage}{0.4\textwidth} %
				\fontsize{9}{14}\selectfont %
				$A_1$(turn right)%
			\end{minipage}
		}
      \put(12.1,35.9){%
			\begin{minipage}{0.4\textwidth} %
				\fontsize{9}{14}\selectfont %
				$A_2$(turn left)%
			\end{minipage}
		}
      \put(10.3,33.6){%
			\begin{minipage}{0.4\textwidth} %
				\fontsize{9}{14}\selectfont %
				$A_3$(drift)%
			\end{minipage}
		}
      \put(48.5,40.5){
			\begin{minipage}{0.4\textwidth} %
				\fontsize{9}{14}\selectfont %
				$A_0$(move forward)%
			\end{minipage}
		}
      \put(47,38.2){%
			\begin{minipage}{0.4\textwidth} %
				\fontsize{9}{14}\selectfont %
				$A_1$(turn right)%
			\end{minipage}
		}
      \put(46.1,35.9){%
			\begin{minipage}{0.4\textwidth} %
				\fontsize{9}{14}\selectfont %
				$A_2$(turn left)%
			\end{minipage}
		}
      \put(44.3,33.6){%
			\begin{minipage}{0.4\textwidth} %
				\fontsize{9}{14}\selectfont %
				$A_3$(drift)%
			\end{minipage}
		}
\end{overpic}
	\caption{(a) Trajectory of the swimmer (red line) with current SAC agent and the force experienced (red arrow) and action choice over $t/T=50,51.25,...,55$ and (b) $t/T=180,181.25,...,185$ in cave exploration task. They correspond to two turnings at the end of the two branches in Figs.~\ref{fig:simple_traj}(a) and (c).}\label{fig:turningComplex} 
\end{figure}

As shown in Fig.~\ref{fig:turningComplex}, the swimmer performs turning maneuvers analogous to the strategy in Fig.~\ref{fig:BigObs}(d).
Panels (a) and (b) capture two such turn-back events during the mission.
This behavior indicates that the swimmer is endowed with the capability to detect frontal obstacles through real-time force and torque sensing, enabling proactive turn initiation. 
Consequently, it not only avoids collisions but also strategically exploits the wall reflection effect to execute efficient turns. 

\begin{figure}
	\centering
	\begin{overpic}[scale=0.4]{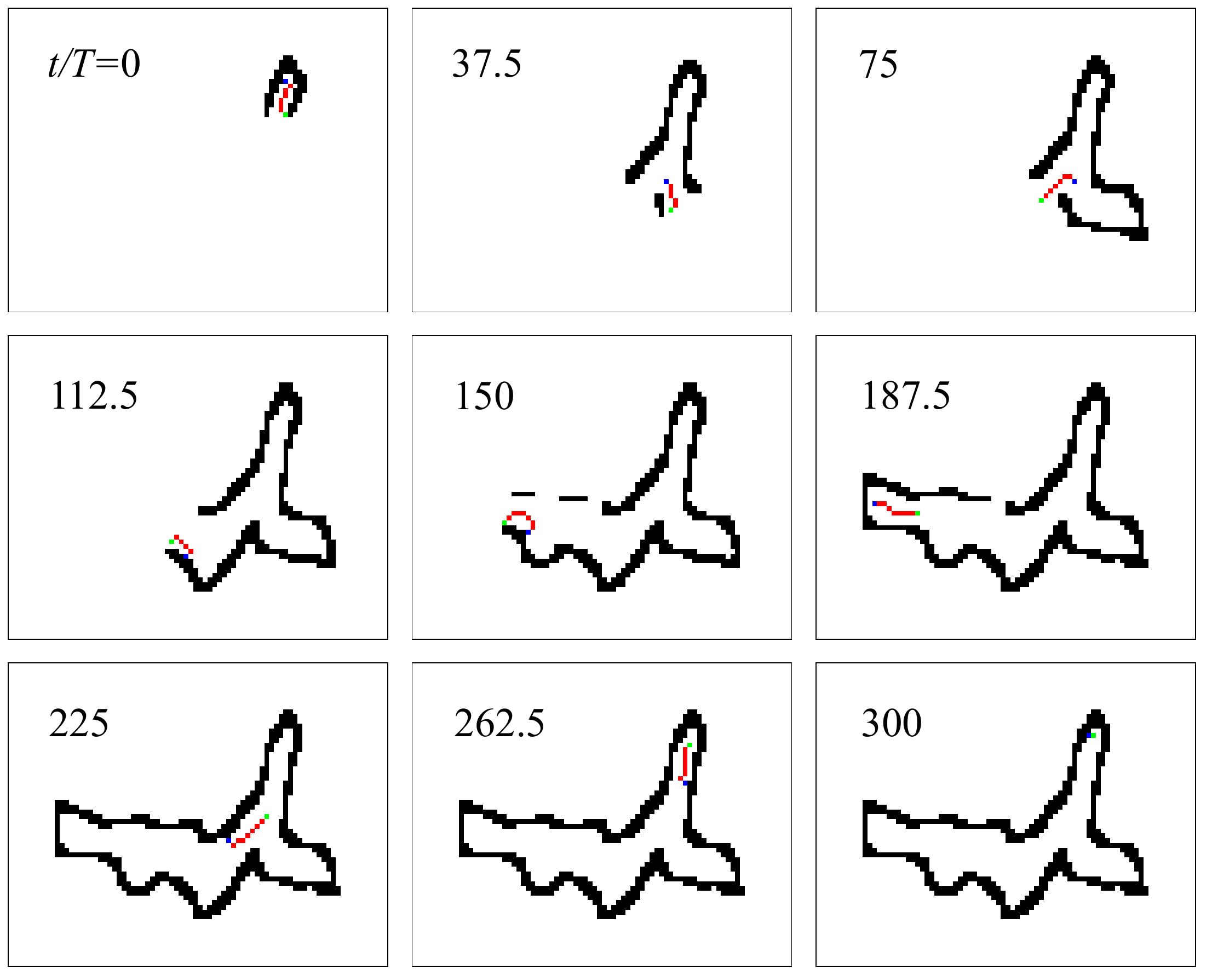}
		\put(29,-3){%
			\begin{minipage}{0.4\textwidth} %
				\fontsize{14}{14}\selectfont %
				$x$ %
			\end{minipage}
		}
      \put(-24.5,39){%
			\begin{minipage}{0.4\textwidth} %
				\fontsize{14}{14}\selectfont %
				$y$ %
			\end{minipage}
		}
\end{overpic}
	\caption{Maze map constructed in real time with position of the swimmer (blue dot), planned path (red dot), pilot and target (both in green dot) at different $t/T$ (marked at the left upper of each panel). The swimmer's task is to explore the closed boundary of the whole cave. Upon finishing the exploration, it navigates to the start point.}\label{fig:simple_Astar} 
\end{figure}

The path planning strategy is depicted in Fig.~\ref{fig:simple_Astar}. 
This strategy involves continuously reassigning the target to the nearest open end within the explored environment. 
In the presented scenario, because the path is straightforward and the target remains in close vicinity to the swimmer, the pilot is set to be identical to the primary target.

\section{\label{sec:conclu}Conclusion}
We present a DRL framework for autonomous navigation of a jellyfish-like swimmer in complex, obstacle-laden fluid environments, specifically addressing the significant influence of vortex-wall interactions on swimming near the wall.  
By augmenting the agent's state representation to explicitly include the real-time forces and torque experienced by the swimmer, we demonstrate how DRL agents perceive and respond to wall boundaries. 
Our analysis reveals that these mechanical feedback signals provide necessary information, which is absent in the previous model \citep{chen2024DRL} with kinematic and geometric state spaces (e.g., position and velocity). 

Especially, vortex-wall interactions induce significant and often non-monotonic variations in the forces and torque acting on the swimmer at moderate to high Reynolds numbers. 
These interactions arise from the reflection and reorganization of the swimmer’s shed vortices by nearby walls, leading to asymmetric pressure distributions and unsteady lateral forces. 
Such hydrodynamic phenomena are captured in real time through the force and torque signals, which serve as a highly sensitive directional proximity sensor. 
For instance, strong frontal drag indicates a head-on obstacle, while lateral forces and torque reveal the presence and orientation of side walls. 
These mechanical cues allow the agent to detect complex near-field flow effects, such as pressure gradients and vortex shedding, that are inaccessible through geometric or kinematic states alone.

This direct physical feedback improves the agent's decision-making.
Compared to previous agent lacking force and torque input, our augmented agent exhibits anticipatory and targeted maneuvers, initiating smoother turns earlier and exploiting wall-induced forces (e.g., leveraging lateral drag for efficient reorientation) rather than resorting to reactive, oscillatory collision-avoidance behaviors.
This leads to enhanced performance in both simple target pursuit with obstacles and complex navigation scenarios, characterized by faster task completion ($40\%$ faster for the single obstacle task).

Our findings underscore the critical role of embodied mechanical interaction for intelligent control in fluid-immersed systems. 
The success of force and torque feedback mirrors principles in robotic tactile perception, where direct contact forces guide dexterous manipulation. 
This work validates the feasibility of using DRL-controlled bio-inspired swimmers for high-risk applications like underwater cave exploration.
More broadly, it establishes force and torque feedback as a vital sensory modality for next-generation autonomous systems operating in confined, fluid-dominated environments. 

Despite its contributions, this study has limitations that suggest directions for future work. 
Notably, our model relies solely on hydrodynamic forces and torques to sense environmental boundaries, whereas natural jellyfish such as Tripedalia cystophora integrates visual cues for obstacle avoidance. 
This discrepancy inspires the development of a multimodal agent by incorporating visual information into the state vector. 
Furthermore, the recent advances in large language model agents show promising potential to surpass traditional neural networks in such complex, embodied scenarios. 
Finally, for real-world exploration, deploying multiple robots concurrently could significantly accelerate data collection.

\begin{acknowledgments}
Numerical simulations were carried out on the TH-2A supercomputer in Guangzhou, China.
This work has been supported by the National Natural Science Foundation of China (Grant Nos.~12525201, 12432010 and 12588201), and the Xplore Prize.
\end{acknowledgments}

\appendix

\section{\label{sec:train_detail}Details on reinforcement learning and training of SAC}
Reinforcement Learning is a type of machine learning where an agent learns to make decisions by interacting with an environment to maximize cumulative rewards. The core principle is based on trial-and-error learning, where the agent takes actions in an environment, observes the outcomes, and adjusts its behavior accordingly. 
The tracking task is formulated as a sequential decision problem solved using reinforcement learning.
The decision-making is modeled as a Markov decision process where the agent makes decision solely based on the current state \citep{richard2018RL}.

The trajectory of the swimmer can be written as
\begin{equation}
	\Gamma_{t}=(s_{0},a_{0},r_{0},...,s_{t-1},a_{t-1},r_{t-1},s_{t},a_{t},r_{t}),
\end{equation}
where $s_{t}$, $a_{t}$, and $r_{t}$ denote the state, the action taken by the agent, and the reward received at a given time $t^*=t\Delta t$, respectively. In the present study, $\Delta t=T/4$, corresponding to the time interval for force application. 
The agent takes action $a_0$ at $t=0$, and then it receives reward $r_0$ and its state changes from $s_0$ to $s_1$. The same process continues until time step $t$.
The action $a_{t}$ taken by the agent is determined solely by the current state $s_{t}$ and the agent's policy $\pi(\Theta, \cdot)$, which is implemented as a neural network with parameters $\Theta$.
The reward function, $r_{t}=r_{t}(s_{t},a_{t})$, depends on the current state and action.

SAC is an offline algorithm where its training does not involve constantly interacting with the environment.
Instead, it learns from history data.
Offline method is preferred in this scenario because of the time cost for simulation of the flow field.
It is very time-consuming to constantly interact with the environment.
The training of SAC is detailed in Algorithm~\ref{alg:SAC}.

\SetKwInOut{REQUIRE}{\textbf{Initialize}}
\begin{algorithm}
\caption{Training of SAC method} 
\label{alg:SAC}
\REQUIRE {load dataset obtained from simulation}
\REQUIRE {policy network $\pi_{p}$ with random parameter $\Theta_{p}$}
\REQUIRE {4 deep Q network $Q_{\phi_1},Q_{\phi_2},Q_{t_1},Q_{t_2}$ with random parameter $\Theta_{\phi_1},\Theta_{\phi_2},\Theta_{t_1}=\Theta_{\phi_1},\Theta_{t_2}=\Theta_{\phi_2}$}
\REQUIRE {hyper parameters learning rate $\varepsilon$ and $\varepsilon^{\prime}$, discount factor $\gamma$, relaxation factor $\tau$, temperature $\alpha$}

\For{$i = 1$ \textbf{to} $N$}{ 
sample $(\boldsymbol{s}_{t},a_{t},r_{t},\boldsymbol{s}_{t+1},D)$ tuple batch from dataset\;
calculate every tuple in the batch \\ 
$\mathbb{E}_a\left(Q_{\phi_1}(s_{t},a)-\alpha \mathrm{log}\left(\pi_p(a|s_{t})\right)\right)=\sum_{i=1}^{N_a} \pi_p(i|s_{t})\left(Q_{\phi_1}(s_{t},i)-\alpha \mathrm{log}(\pi_p(i|s_{t}))\right)$\;
$\Theta_{p}\leftarrow\Theta_{p}+\varepsilon\nabla_p \mathbb{E}_a$\;
$Q_{\mathrm{predict_{1}}}=Q_{\phi_{1}}(s_{t},a_t),Q_{\mathrm{predict_{2}}}=Q_{\phi_{2}}(s_{t},a_t)$\;
$Q^\prime = \mathrm{min}\left(\sum_{i=1}^{N_a}Q_{t_1}(s_{t+1},i)\pi_p(i|s_{t+1}),\sum_{i=1}^{N_a}Q_{t_2}(s_{t+1},i)\pi_p(i|s_{t+1})\right)$\;
$Q_{\mathrm{target}}=r_{t}+\gamma(1-D)\left(Q^\prime
- \alpha\sum_{i=1}^{N_a}\pi_p(i|s_{t+1})\mathrm{log}(\pi_p(i|s_{t+1})) \right)$\;
$L_1 = \frac{1}{2}(Q_{\mathrm{predict_{1}}}-Q_{\mathrm{target}})^2, L_2 = \frac{1}{2}(Q_{\mathrm{predict_{2}}}-Q_{\mathrm{target}})^2$\;
$\Theta_{\phi_1}\leftarrow\Theta_{\phi_1}-\varepsilon^\prime \nabla_{\phi_1}L_1$\;
$\Theta_{\phi_2}\leftarrow\Theta_{\phi_2}-\varepsilon^\prime \nabla_{\phi_2}L_2$\;
$\Theta_{t_1}\leftarrow (1-\tau)\Theta_{t_1}+\tau\Theta_{\phi_1}$\;
$\Theta_{t_2}\leftarrow (1-\tau)\Theta_{t_2}+\tau\Theta_{\phi_2}$\;
}
\end{algorithm}

\section{\label{sec:Astar}A* algorithm for navigating}
 A* (A-star) algorithm is a widely used pathfinding and graph traversal algorithm \citep{Peter1968Astar}. 
 It efficiently finds the shortest path between nodes in a graph by combining the actual cost from the start node with an estimated heuristic cost to the goal. This heuristic approach allows A* to explore fewer paths compared to other algorithms like Dijkstra’s, making it both optimal and complete when using an admissible heuristic. A* is commonly applied in fields such as robotics, video games, and navigation systems for route planning.
For the navigation in present study, obstacles are represented by Lagrangian points of their boundaries. 
The domain with obstacles is discretized into a grid maze where the A* algorithm is applicable.

The algorithm is detailed in Algorithm~\ref{alg:Astar}.
The cost of a position $P$ in the algorithm is a heuristic function
\begin{equation}
	\mathrm{cost}(P) = n(P) + \lVert \mathcal{P}-\mathcal{E} \rVert_{1} + d(P),
\end{equation}
where the first term $n(P)$ is the path length from $\mathcal{S}$ to $P$,
$\lVert \mathcal{P}-\mathcal{E} \rVert_{1}$ is the sum of coordinate difference between $P$ and $\mathcal{E}$ and $d(P)$ the distance from the closest obstacle if there is obstacle point in range.
This heuristic function encourages the path to be short and keep distance to obstacle at the same time.

\begin{algorithm}
\caption{A* algorithm} 
\label{alg:Astar}
\REQUIRE {empty queue $\mathcal{Q}$, start position $\mathcal{S}$, end position $\mathcal{E}$}
push $\mathcal{S}$ to $\mathcal{Q}$\;
\While{$\mathcal{Q}$ is not empty}{
    $x \gets \text{pop element with smallest cost from } \mathcal{Q}$\;
     \If{$x$ is $\mathcal{E}$}{
        backtrack from $\mathcal{E}$ \;
        return the trajectory from $\mathcal{S}$ to $\mathcal{E}$\;
    }
    push non-repetitive neighbor positions of $x$ to $\mathcal{Q}$\;
}
\end{algorithm}

Figures~\ref{fig:outflow_Astar} and \ref{fig:simple_Astar} showcase the process of environment perception and path planning of the unknown environment in Section~\ref{sec:obstacle}.
The path consists of the grid points connecting the grid point that is closest to the swimmer to the target grid point.
After the path is established, the pilot is set to the grid point on the path where there is no obstacles between it and the swimmer and its distance to the swimmer is closest to $\mathcal{R}$.
One exception is that if the swimmer is close enough to the target where its distance to the target is smaller than $\mathcal{R}$, and there is no obstacle between swimmer and the target, the pilot is set to the target point.
In the scenario depicted in Fig.~\ref{fig:outflow_Astar}, the target is known and fixed on the maze map, with the mission being to reach this target. In contrast, for the scenario in Fig.~\ref{fig:simple_Astar}, no target is predefined; instead, the objective is to fully explore the boundary of the domain. Accordingly, the target is computed in real time based on the environment sensed by the swimmer. The selected target is the farthest reachable point detected within the explored region. The A* algorithm is then employed to plan a path from the swimmer's current position to this target, while a pilot is also calculated as described previously.

\nocite{*}

\bibliography{apssamp}

\end{document}